\hsize=31pc
\vsize=49pc
\lineskip=0pt
\parskip=0pt plus 1pt
\hfuzz=1pt  
\vfuzz=2pt
\pretolerance=2500
\tolerance=5000
\vbadness=5000
\hbadness=5000
\widowpenalty=500
\clubpenalty=200
\brokenpenalty=500
\predisplaypenalty=200
\voffset=-1pc
\nopagenumbers     
\catcode`@=11
\newif\ifams
\amstrue
%
%%%%%%%%%%%%%%%%%%%%%%%%%%%%%%%%%%%%%%%%%%%%%%%%%%%%%%%%%%%%%
%                                                           %
%  The following section may be commented out and           %
%  \ifams set to either \amstrue to use the AMS fonts       %
%  or \amsfalse if they are not available                   %
%                                                           %
%%%%%%%%%%%%%%%%%%%%%%%%%%%%%%%%%%%%%%%%%%%%%%%%%%%%%%%%%%%%%
%
%\def\Yesreply{Y }
%\def\Noreply{N }
%\def\yesreply{y }
%\def\noreply{n }
%\newif\ifnotyorn
%\message{Do you want to use AMSfonts, msam and msbm? Y or N: }%
%\loop
%\read-1 to \reply
%\ifx\reply\yesreply\global\amstrue\notyornfalse
%\else\ifx\reply\Yesreply\global\amstrue\notyornfalse
%\else\ifx\reply\noreply\global\amsfalse\notyornfalse
%\else\ifx\reply\Noreply\global\amsfalse\notyornfalse
%\else\notyorntrue
%\message{Please type y or Y  (Yes) or n or N (No)}\fi\fi\fi\fi
%\ifnotyorn\repeat
%
%%%%%%%%%%%%%%%%%%%%%%%%%%%%%%%%%%%%%%%%%%%%%%%%%%%%%%%%%%%%
%
\newfam\bdifam
\newfam\bsyfam
\newfam\bssfam
\newfam\msafam
\newfam\msbfam
\newif\ifxxpt   
\newif\ifxviipt 
\newif\ifxivpt  
\newif\ifxiipt  
\newif\ifxipt   
\newif\ifxpt    
\newif\ifixpt   
\newif\ifviiipt 
\newif\ifviipt  
\newif\ifvipt   
\newif\ifvpt    
%
% Headings in 20pt, 17pt or 14pt
%
\def\headsize#1#2{\def\headb@seline{#2}%
                \ifnum#1=20\def\HEAD{twenty}%
                           \def\smHEAD{twelve}%
                           \def\vsHEAD{nine}%
                           \ifxxpt\else\xdef\f@ntsize{\HEAD}%
                           \def\m@g{4}\def\s@ze{20.74}%
                           \loadheadfonts\xxpttrue\fi
                           \ifxiipt\else\xdef\f@ntsize{\smHEAD}%
                           \def\m@g{1}\def\s@ze{12}%
                           \loadxiiptfonts\xiipttrue\fi
                           \ifixpt\else\xdef\f@ntsize{\vsHEAD}%
                           \def\s@ze{9}%
                           \loadsmallfonts\ixpttrue\fi
                      \else
                \ifnum#1=17\def\HEAD{seventeen}%
                           \def\smHEAD{eleven}%
                           \def\vsHEAD{eight}%
                           \ifxviipt\else\xdef\f@ntsize{\HEAD}%
                           \def\m@g{3}\def\s@ze{17.28}%
                           \loadheadfonts\xviipttrue\fi
                           \ifxipt\else\xdef\f@ntsize{\smHEAD}%
                           \loadxiptfonts\xipttrue\fi
                           \ifviiipt\else\xdef\f@ntsize{\vsHEAD}%
                           \def\s@ze{8}%
                           \loadsmallfonts\viiipttrue\fi
                      \else\def\HEAD{fourteen}%
                           \def\smHEAD{ten}%
                           \def\vsHEAD{seven}%
                           \ifxivpt\else\xdef\f@ntsize{\HEAD}%
                           \def\m@g{2}\def\s@ze{14.4}%
                           \loadheadfonts\xivpttrue\fi
                           \ifxpt\else\xdef\f@ntsize{\smHEAD}%
                           \def\s@ze{10}%
                           \loadxptfonts\xpttrue\fi
                           \ifviipt\else\xdef\f@ntsize{\vsHEAD}%
                           \def\s@ze{7}%
                           \loadviiptfonts\viipttrue\fi
                \ifnum#1=14\else
                \message{Header size should be 20, 17 or 14 point
                              will now default to 14pt}\fi
                \fi\fi\headfonts}
%
% Text in 12pt, 11pt or 10pt 
%
\def\textsize#1#2{\def\textb@seline{#2}%
                 \ifnum#1=12\def\TEXT{twelve}%
                           \def\smTEXT{eight}%
                           \def\vsTEXT{six}%
                           \ifxiipt\else\xdef\f@ntsize{\TEXT}%
                           \def\m@g{1}\def\s@ze{12}%
                           \loadxiiptfonts\xiipttrue\fi
                           \ifviiipt\else\xdef\f@ntsize{\smTEXT}%
                           \def\s@ze{8}%
                           \loadsmallfonts\viiipttrue\fi
                           \ifvipt\else\xdef\f@ntsize{\vsTEXT}%
                           \def\s@ze{6}%
                           \loadviptfonts\vipttrue\fi
                      \else
                \ifnum#1=11\def\TEXT{eleven}%
                           \def\smTEXT{seven}%
                           \def\vsTEXT{five}%
                           \ifxipt\else\xdef\f@ntsize{\TEXT}%
                           \def\s@ze{11}%
                           \loadxiptfonts\xipttrue\fi
                           \ifviipt\else\xdef\f@ntsize{\smTEXT}%
                           \loadviiptfonts\viipttrue\fi
                           \ifvpt\else\xdef\f@ntsize{\vsTEXT}%
                           \def\s@ze{5}%
                           \loadvptfonts\vpttrue\fi
                      \else\def\TEXT{ten}%
                           \def\smTEXT{seven}%
                           \def\vsTEXT{five}%
                           \ifxpt\else\xdef\f@ntsize{\TEXT}%
                           \loadxptfonts\xpttrue\fi
                           \ifviipt\else\xdef\f@ntsize{\smTEXT}%
                           \def\s@ze{7}%
                           \loadviiptfonts\viipttrue\fi
                           \ifvpt\else\xdef\f@ntsize{\vsTEXT}%
                           \def\s@ze{5}%
                           \loadvptfonts\vpttrue\fi
                \ifnum#1=10\else
                \message{Text size should be 12, 11 or 10 point
                              will now default to 10pt}\fi
                \fi\fi\textfonts}
%
% Small sized material in 10pt, 9pt or 8pt
%
\def\smallsize#1#2{\def\smallb@seline{#2}%
                 \ifnum#1=10\def\SMALL{ten}%
                           \def\smSMALL{seven}%
                           \def\vsSMALL{five}%
                           \ifxpt\else\xdef\f@ntsize{\SMALL}%
                           \loadxptfonts\xpttrue\fi
                           \ifviipt\else\xdef\f@ntsize{\smSMALL}%
                           \def\s@ze{7}%
                           \loadviiptfonts\viipttrue\fi
                           \ifvpt\else\xdef\f@ntsize{\vsSMALL}%
                           \def\s@ze{5}%
                           \loadvptfonts\vpttrue\fi
                       \else
                 \ifnum#1=9\def\SMALL{nine}%
                           \def\smSMALL{six}%
                           \def\vsSMALL{five}%
                           \ifixpt\else\xdef\f@ntsize{\SMALL}%
                           \def\s@ze{9}%
                           \loadsmallfonts\ixpttrue\fi
                           \ifvipt\else\xdef\f@ntsize{\smSMALL}%
                           \def\s@ze{6}%
                           \loadviptfonts\vipttrue\fi
                           \ifvpt\else\xdef\f@ntsize{\vsSMALL}%
                           \def\s@ze{5}%
                           \loadvptfonts\vpttrue\fi
                       \else
                           \def\SMALL{eight}%
                           \def\smSMALL{six}%
                           \def\vsSMALL{five}%
                           \ifviiipt\else\xdef\f@ntsize{\SMALL}%
                           \def\s@ze{8}%
                           \loadsmallfonts\viiipttrue\fi
                           \ifvipt\else\xdef\f@ntsize{\smSMALL}%
                           \def\s@ze{6}%
                           \loadviptfonts\vipttrue\fi
                           \ifvpt\else\xdef\f@ntsize{\vsSMALL}%
                           \def\s@ze{5}%
                           \loadvptfonts\vpttrue\fi
                 \ifnum#1=8\else\message{Small size should be 10, 9 or 
                            8 point will now default to 8pt}\fi
                \fi\fi\smallfonts}
\def\F@nt{\expandafter\font\csname}
\def\Sk@w{\expandafter\skewchar\csname}
\def\@nd{\endcsname}
\def\@step#1{ scaled \magstep#1}
\def\@half{ scaled \magstephalf}
\def\@t#1{ at #1pt}
%
% For 14, 17 and 20 point fonts use \loadheadfonts
%
\def\loadheadfonts{\bigf@nts
\F@nt \f@ntsize bdi\@nd=cmmib10 \@t{\s@ze}%
\Sk@w \f@ntsize bdi\@nd='177
\F@nt \f@ntsize bsy\@nd=cmbsy10 \@t{\s@ze}%
\Sk@w \f@ntsize bsy\@nd='60
\F@nt \f@ntsize bss\@nd=cmssbx10 \@t{\s@ze}}
%
% For 12 point fonts use \loadxiiptfonts
%
\def\loadxiiptfonts{\bigf@nts
\F@nt \f@ntsize bdi\@nd=cmmib10 \@step{\m@g}%
\Sk@w \f@ntsize bdi\@nd='177
\F@nt \f@ntsize bsy\@nd=cmbsy10 \@step{\m@g}%
\Sk@w \f@ntsize bsy\@nd='60
\F@nt \f@ntsize bss\@nd=cmssbx10 \@step{\m@g}}
%
% For 11 point fonts use \loadxiptfonts
%
\def\loadxiptfonts{%
\font\elevenrm=cmr10 \@half
\font\eleveni=cmmi10 \@half
\skewchar\eleveni='177
\font\elevensy=cmsy10 \@half
\skewchar\elevensy='60
\font\elevenex=cmex10 \@half
\font\elevenit=cmti10 \@half
\font\elevensl=cmsl10 \@half
\font\elevenbf=cmbx10 \@half
\font\eleventt=cmtt10 \@half
\ifams\font\elevenmsa=msam10 \@half
\font\elevenmsb=msbm10 \@half\else\fi
\font\elevenbdi=cmmib10 \@half
\skewchar\elevenbdi='177
\font\elevenbsy=cmbsy10 \@half
\skewchar\elevenbsy='60
\font\elevenbss=cmssbx10 \@half}
%
% For 10 point fonts use \loadxptfonts
%
\def\loadxptfonts{%
\font\tenbdi=cmmib10
\skewchar\tenbdi='177
\font\tenbsy=cmbsy10 
\skewchar\tenbsy='60
\ifams\font\tenmsa=msam10 
\font\tenmsb=msbm10\else\fi
\font\tenbss=cmssbx10}% 
%
% For 8 and 9 point fonts use \loadsmallfonts
%
\def\loadsmallfonts{\smallf@nts
\ifams
\F@nt \f@ntsize ex\@nd=cmex\s@ze
\else
\F@nt \f@ntsize ex\@nd=cmex10\fi
\F@nt \f@ntsize it\@nd=cmti\s@ze
\F@nt \f@ntsize sl\@nd=cmsl\s@ze
\F@nt \f@ntsize tt\@nd=cmtt\s@ze}
%
% For 7 point fonts use \loadviiptfonts
%
\def\loadviiptfonts{%
\font\sevenit=cmti7
\font\sevensl=cmsl8 at 7pt
\ifams\font\sevenmsa=msam7 
\font\sevenmsb=msbm7
\font\sevenex=cmex7
\font\sevenbsy=cmbsy7
\font\sevenbdi=cmmib7\else
\font\sevenex=cmex10
\font\sevenbsy=cmbsy10 at 7pt
\font\sevenbdi=cmmib10 at 7pt\fi
\skewchar\sevenbsy='60
\skewchar\sevenbdi='177
\font\sevenbss=cmssbx10 at 7pt}% 
%
%  For 6 point fonts use \loadviptfonts
%
\def\loadviptfonts{\smallf@nts
\ifams\font\sixex=cmex7 at 6pt\else
\font\sixex=cmex10\fi
\font\sixit=cmti7 at 6pt}
%
% For 5 point fonts use \loadvptfonts
%
\def\loadvptfonts{%
\font\fiveit=cmti7 at 5pt
\ifams\font\fiveex=cmex7 at 5pt
\font\fivebdi=cmmib5
\font\fivebsy=cmbsy5
\font\fivemsa=msam5 
\font\fivemsb=msbm5\else
\font\fiveex=cmex10
\font\fivebdi=cmmib10 at 5pt
\font\fivebsy=cmbsy10 at 5pt\fi
\skewchar\fivebdi='177
\skewchar\fivebsy='60
\font\fivebss=cmssbx10 at 5pt}
\def\bigf@nts{%
\F@nt \f@ntsize rm\@nd=cmr10 \@step{\m@g}%
\F@nt \f@ntsize i\@nd=cmmi10 \@step{\m@g}%
\Sk@w \f@ntsize i\@nd='177
\F@nt \f@ntsize sy\@nd=cmsy10 \@step{\m@g}%
\Sk@w \f@ntsize sy\@nd='60
\F@nt \f@ntsize ex\@nd=cmex10 \@step{\m@g}%
\F@nt \f@ntsize it\@nd=cmti10 \@step{\m@g}%
\F@nt \f@ntsize sl\@nd=cmsl10 \@step{\m@g}%
\F@nt \f@ntsize bf\@nd=cmbx10 \@step{\m@g}%
\F@nt \f@ntsize tt\@nd=cmtt10 \@step{\m@g}%
\ifams
\F@nt \f@ntsize msa\@nd=msam10 \@step{\m@g}%
\F@nt \f@ntsize msb\@nd=msbm10 \@step{\m@g}\else\fi}
\def\smallf@nts{%
\F@nt \f@ntsize rm\@nd=cmr\s@ze
\F@nt \f@ntsize i\@nd=cmmi\s@ze 
\Sk@w \f@ntsize i\@nd='177
\F@nt \f@ntsize sy\@nd=cmsy\s@ze
\Sk@w \f@ntsize sy\@nd='60
\F@nt \f@ntsize bf\@nd=cmbx\s@ze 
\ifams
\F@nt \f@ntsize bdi\@nd=cmmib\s@ze 
\F@nt \f@ntsize bsy\@nd=cmbsy\s@ze 
\F@nt \f@ntsize msa\@nd=msam\s@ze 
\F@nt \f@ntsize msb\@nd=msbm\s@ze
\else
\F@nt \f@ntsize bdi\@nd=cmmib10 \@t{\s@ze}% 
\F@nt \f@ntsize bsy\@nd=cmbsy10 \@t{\s@ze}\fi 
\Sk@w \f@ntsize bdi\@nd='177
\Sk@w \f@ntsize bsy\@nd='60
\F@nt \f@ntsize bss\@nd=cmssbx10 \@t{\s@ze}}% 
%
% Fonts for headings 
%
\def\headfonts{%
\textfont0=\csname\HEAD rm\@nd        
\scriptfont0=\csname\smHEAD rm\@nd
\scriptscriptfont0=\csname\vsHEAD rm\@nd
\def\rm{\fam0\csname\HEAD rm\@nd
\def\sc{\csname\smHEAD rm\@nd}}%
\textfont1=\csname\HEAD i\@nd         
\scriptfont1=\csname\smHEAD i\@nd
\scriptscriptfont1=\csname\vsHEAD i\@nd
\textfont2=\csname\HEAD sy\@nd        
\scriptfont2=\csname\smHEAD sy\@nd
\scriptscriptfont2=\csname\vsHEAD sy\@nd
\textfont3=\csname\HEAD ex\@nd        
\scriptfont3=\csname\smHEAD ex\@nd
\scriptscriptfont3=\csname\smHEAD ex\@nd
\textfont\itfam=\csname\HEAD it\@nd   
\scriptfont\itfam=\csname\smHEAD it\@nd
\scriptscriptfont\itfam=\csname\vsHEAD it\@nd
\def\it{\fam\itfam\csname\HEAD it\@nd
\def\sc{\csname\smHEAD it\@nd}}%
\textfont\slfam=\csname\HEAD sl\@nd   
\def\sl{\fam\slfam\csname\HEAD sl\@nd
\def\sc{\csname\smHEAD sl\@nd}}%
\textfont\bffam=\csname\HEAD bf\@nd   
\scriptfont\bffam=\csname\smHEAD bf\@nd
\scriptscriptfont\bffam=\csname\vsHEAD bf\@nd
\def\bf{\fam\bffam\csname\HEAD bf\@nd
\def\sc{\csname\smHEAD bf\@nd}}%
\textfont\ttfam=\csname\HEAD tt\@nd   
\def\tt{\fam\ttfam\csname\HEAD tt\@nd}%
\textfont\bdifam=\csname\HEAD bdi\@nd 
\scriptfont\bdifam=\csname\smHEAD bdi\@nd
\scriptscriptfont\bdifam=\csname\vsHEAD bdi\@nd
\def\bdi{\fam\bdifam\csname\HEAD bdi\@nd}%
\textfont\bsyfam=\csname\HEAD bsy\@nd 
\scriptfont\bsyfam=\csname\smHEAD bsy\@nd
\def\bsy{\fam\bsyfam\csname\HEAD bsy\@nd}%
\textfont\bssfam=\csname\HEAD bss\@nd 
\scriptfont\bssfam=\csname\smHEAD bss\@nd
\scriptscriptfont\bssfam=\csname\vsHEAD bss\@nd
\def\bss{\fam\bssfam\csname\HEAD bss\@nd}%
\ifams
\textfont\msafam=\csname\HEAD msa\@nd 
\scriptfont\msafam=\csname\smHEAD msa\@nd
\scriptscriptfont\msafam=\csname\vsHEAD msa\@nd
\textfont\msbfam=\csname\HEAD msb\@nd 
\scriptfont\msbfam=\csname\smHEAD msb\@nd
\scriptscriptfont\msbfam=\csname\vsHEAD msb\@nd
\else\fi
\normalbaselineskip=\headb@seline pt%
\setbox\strutbox=\hbox{\vrule height.7\normalbaselineskip 
depth.3\baselineskip width0pt}%
\def\sc{\csname\smHEAD rm\@nd}\normalbaselines\bf}
%
% Fonts for text
%
\def\textfonts{%
\textfont0=\csname\TEXT rm\@nd        
\scriptfont0=\csname\smTEXT rm\@nd
\scriptscriptfont0=\csname\vsTEXT rm\@nd
\def\rm{\fam0\csname\TEXT rm\@nd
\def\sc{\csname\smTEXT rm\@nd}}%
\textfont1=\csname\TEXT i\@nd         
\scriptfont1=\csname\smTEXT i\@nd
\scriptscriptfont1=\csname\vsTEXT i\@nd
\textfont2=\csname\TEXT sy\@nd        
\scriptfont2=\csname\smTEXT sy\@nd
\scriptscriptfont2=\csname\vsTEXT sy\@nd
\textfont3=\csname\TEXT ex\@nd        
\scriptfont3=\csname\smTEXT ex\@nd
\scriptscriptfont3=\csname\smTEXT ex\@nd
\textfont\itfam=\csname\TEXT it\@nd   
\scriptfont\itfam=\csname\smTEXT it\@nd
\scriptscriptfont\itfam=\csname\vsTEXT it\@nd
\def\it{\fam\itfam\csname\TEXT it\@nd
\def\sc{\csname\smTEXT it\@nd}}%
\textfont\slfam=\csname\TEXT sl\@nd   
\def\sl{\fam\slfam\csname\TEXT sl\@nd
\def\sc{\csname\smTEXT sl\@nd}}%
\textfont\bffam=\csname\TEXT bf\@nd   
\scriptfont\bffam=\csname\smTEXT bf\@nd
\scriptscriptfont\bffam=\csname\vsTEXT bf\@nd
\def\bf{\fam\bffam\csname\TEXT bf\@nd
\def\sc{\csname\smTEXT bf\@nd}}%
\textfont\ttfam=\csname\TEXT tt\@nd   
\def\tt{\fam\ttfam\csname\TEXT tt\@nd}%
\textfont\bdifam=\csname\TEXT bdi\@nd 
\scriptfont\bdifam=\csname\smTEXT bdi\@nd
\scriptscriptfont\bdifam=\csname\vsTEXT bdi\@nd
\def\bdi{\fam\bdifam\csname\TEXT bdi\@nd}%
\textfont\bsyfam=\csname\TEXT bsy\@nd 
\scriptfont\bsyfam=\csname\smTEXT bsy\@nd
\def\bsy{\fam\bsyfam\csname\TEXT bsy\@nd}%
\textfont\bssfam=\csname\TEXT bss\@nd 
\scriptfont\bssfam=\csname\smTEXT bss\@nd
\scriptscriptfont\bssfam=\csname\vsTEXT bss\@nd
\def\bss{\fam\bssfam\csname\TEXT bss\@nd}%
\ifams
\textfont\msafam=\csname\TEXT msa\@nd 
\scriptfont\msafam=\csname\smTEXT msa\@nd
\scriptscriptfont\msafam=\csname\vsTEXT msa\@nd
\textfont\msbfam=\csname\TEXT msb\@nd 
\scriptfont\msbfam=\csname\smTEXT msb\@nd
\scriptscriptfont\msbfam=\csname\vsTEXT msb\@nd
\else\fi
\normalbaselineskip=\textb@seline pt
\setbox\strutbox=\hbox{\vrule height.7\normalbaselineskip 
depth.3\baselineskip width0pt}%
\everymath{}%
\def\sc{\csname\smTEXT rm\@nd}\normalbaselines\rm}
%
% Fonts for small material (captions, footnotes etc)
%
\def\smallfonts{%
\textfont0=\csname\SMALL rm\@nd        
\scriptfont0=\csname\smSMALL rm\@nd
\scriptscriptfont0=\csname\vsSMALL rm\@nd
\def\rm{\fam0\csname\SMALL rm\@nd
\def\sc{\csname\smSMALL rm\@nd}}%
\textfont1=\csname\SMALL i\@nd         
\scriptfont1=\csname\smSMALL i\@nd
\scriptscriptfont1=\csname\vsSMALL i\@nd
\textfont2=\csname\SMALL sy\@nd        
\scriptfont2=\csname\smSMALL sy\@nd
\scriptscriptfont2=\csname\vsSMALL sy\@nd
\textfont3=\csname\SMALL ex\@nd        
\scriptfont3=\csname\smSMALL ex\@nd
\scriptscriptfont3=\csname\smSMALL ex\@nd
\textfont\itfam=\csname\SMALL it\@nd   
\scriptfont\itfam=\csname\smSMALL it\@nd
\scriptscriptfont\itfam=\csname\vsSMALL it\@nd
\def\it{\fam\itfam\csname\SMALL it\@nd
\def\sc{\csname\smSMALL it\@nd}}%
\textfont\slfam=\csname\SMALL sl\@nd   
\def\sl{\fam\slfam\csname\SMALL sl\@nd
\def\sc{\csname\smSMALL sl\@nd}}%
\textfont\bffam=\csname\SMALL bf\@nd   
\scriptfont\bffam=\csname\smSMALL bf\@nd
\scriptscriptfont\bffam=\csname\vsSMALL bf\@nd
\def\bf{\fam\bffam\csname\SMALL bf\@nd
\def\sc{\csname\smSMALL bf\@nd}}%
\textfont\ttfam=\csname\SMALL tt\@nd   
\def\tt{\fam\ttfam\csname\SMALL tt\@nd}%
\textfont\bdifam=\csname\SMALL bdi\@nd 
\scriptfont\bdifam=\csname\smSMALL bdi\@nd
\scriptscriptfont\bdifam=\csname\vsSMALL bdi\@nd
\def\bdi{\fam\bdifam\csname\SMALL bdi\@nd}%
\textfont\bsyfam=\csname\SMALL bsy\@nd 
\scriptfont\bsyfam=\csname\smSMALL bsy\@nd
\def\bsy{\fam\bsyfam\csname\SMALL bsy\@nd}%
\textfont\bssfam=\csname\SMALL bss\@nd 
\scriptfont\bssfam=\csname\smSMALL bss\@nd
\scriptscriptfont\bssfam=\csname\vsSMALL bss\@nd
\def\bss{\fam\bssfam\csname\SMALL bss\@nd}%
\ifams
\textfont\msafam=\csname\SMALL msa\@nd 
\scriptfont\msafam=\csname\smSMALL msa\@nd
\scriptscriptfont\msafam=\csname\vsSMALL msa\@nd
\textfont\msbfam=\csname\SMALL msb\@nd 
\scriptfont\msbfam=\csname\smSMALL msb\@nd
\scriptscriptfont\msbfam=\csname\vsSMALL msb\@nd
\else\fi
\normalbaselineskip=\smallb@seline pt%
\setbox\strutbox=\hbox{\vrule height.7\normalbaselineskip 
depth.3\baselineskip width0pt}%
\everymath{}%
\def\sc{\csname\smSMALL rm\@nd}\normalbaselines\rm}%
\everydisplay{\indenteddisplay
   \gdef\labeltype{\eqlabel}}%
%
%%%%%%%%%%%%%%%%%%%%%%%%%%%%%%%%%%%%%%%%%%%%%%%%%%%%%%%%%%%
%                                                         %
%  Macros to define extra maths symbols                   %
%                                                         %
%%%%%%%%%%%%%%%%%%%%%%%%%%%%%%%%%%%%%%%%%%%%%%%%%%%%%%%%%%%
%
\def\hexnumber@#1{\ifcase#1 0\or 1\or 2\or 3\or 4\or 5\or 6\or 7\or 8\or
 9\or A\or B\or C\or D\or E\or F\fi}
\edef\bffam@{\hexnumber@\bffam}
\edef\bdifam@{\hexnumber@\bdifam}
\edef\bsyfam@{\hexnumber@\bsyfam}
\def\undefine#1{\let#1\undefined}
\def\newsymbol#1#2#3#4#5{\let\next@\relax
 \ifnum#2=\thr@@\let\next@\bdifam@\else
 \ifams
 \ifnum#2=\@ne\let\next@\msafam@\else
 \ifnum#2=\tw@\let\next@\msbfam@\fi\fi
 \fi\fi
 \mathchardef#1="#3\next@#4#5}
\def\mathhexbox@#1#2#3{\relax
 \ifmmode\mathpalette{}{\m@th\mathchar"#1#2#3}%
 \else\leavevmode\hbox{$\m@th\mathchar"#1#2#3$}\fi}

\def\bi#1{{\fam\bdifam\relax#1}}
%
% If file amsmacro is not in current directory
% or somewhere with set path add path before
% file name in following line
%
\ifams\amsfalse       %input amsmacro\fi
%
% Bold italic Greek characters
%
\newsymbol\bitGamma 3000
\newsymbol\bitDelta 3001
\newsymbol\bitTheta 3002
\newsymbol\bitLambda 3003
\newsymbol\bitXi 3004
\newsymbol\bitPi 3005
\newsymbol\bitSigma 3006
\newsymbol\bitUpsilon 3007
\newsymbol\bitPhi 3008
\newsymbol\bitPsi 3009
\newsymbol\bitOmega 300A
\newsymbol\balpha 300B
\newsymbol\bbeta 300C
\newsymbol\bgamma 300D
\newsymbol\bdelta 300E
\newsymbol\bepsilon 300F
\newsymbol\bzeta 3010
\newsymbol\bfeta 3011
\newsymbol\btheta 3012
\newsymbol\biota 3013
\newsymbol\bkappa 3014
\newsymbol\blambda 3015
\newsymbol\bmu 3016
\newsymbol\bnu 3017
\newsymbol\bxi 3018
\newsymbol\bpi 3019
\newsymbol\brho 301A
\newsymbol\bsigma 301B
\newsymbol\btau 301C
\newsymbol\bupsilon 301D
\newsymbol\bphi 301E
\newsymbol\bchi 301F
\newsymbol\bpsi 3020
\newsymbol\bomega 3021
\newsymbol\bvarepsilon 3022
\newsymbol\bvartheta 3023
\newsymbol\bvaromega 3024
\newsymbol\bvarrho 3025
\newsymbol\bvarzeta 3026
\newsymbol\bvarphi 3027
\newsymbol\bpartial 3040
\newsymbol\bell 3060
\newsymbol\bimath 307B
\newsymbol\bjmath 307C
\mathchardef\binfty "0\bsyfam@31
\mathchardef\bnabla "0\bsyfam@72
\mathchardef\bdot "2\bsyfam@01
\mathchardef\bGamma "0\bffam@00
\mathchardef\bDelta "0\bffam@01
\mathchardef\bTheta "0\bffam@02
\mathchardef\bLambda "0\bffam@03
\mathchardef\bXi "0\bffam@04
\mathchardef\bPi "0\bffam@05
\mathchardef\bSigma "0\bffam@06
\mathchardef\bUpsilon "0\bffam@07
\mathchardef\bPhi "0\bffam@08
\mathchardef\bPsi "0\bffam@09
\mathchardef\bOmega "0\bffam@0A
\mathchardef\itGamma "0100
\mathchardef\itDelta "0101
\mathchardef\itTheta "0102
\mathchardef\itLambda "0103
\mathchardef\itXi "0104
\mathchardef\itPi "0105
\mathchardef\itSigma "0106
\mathchardef\itUpsilon "0107
\mathchardef\itPhi "0108
\mathchardef\itPsi "0109
\mathchardef\itOmega "010A
\mathchardef\Gamma "0000
\mathchardef\Delta "0001
\mathchardef\Theta "0002
\mathchardef\Lambda "0003
\mathchardef\Xi "0004
\mathchardef\Pi "0005
\mathchardef\Sigma "0006
\mathchardef\Upsilon "0007
\mathchardef\Phi "0008
\mathchardef\Psi "0009
\mathchardef\Omega "000A
%
% Counter definitions
%
\newcount\firstpage  \firstpage=1  % start page no
\newcount\jnl                      % journal no
\newcount\secno                    % section number
\newcount\subno                    % number of subsection
\newcount\subsubno                 % number of subsubsection
\newcount\appno                    % appendix number
\newcount\tabno                    % table number
\newcount\figno                    % figure number
\newcount\countno                  % equation numbers
\newcount\refno                    % reference number
\newcount\eqlett     \eqlett=97    % equation letter
\newif\ifletter
\newif\ifwide
\newif\ifnotfull
\newif\ifaligned
\newif\ifnumbysec  
\newif\ifappendix
\newif\ifnumapp
\newif\ifssf
\newif\ifppt
\newdimen\t@bwidth
\newdimen\c@pwidth
\newdimen\digitwidth                    %character width
\newdimen\argwidth                      %argument width
\newdimen\secindent    \secindent=5pc   %indentation of maths 
\newdimen\textind    \textind=16pt      %indentation of text
\newdimen\tempval                       %temporary value
\newskip\beforesecskip
\def\beforesecspace{\vskip\beforesecskip\relax}
\newskip\beforesubskip
\def\beforesubspace{\vskip\beforesubskip\relax}
\newskip\beforesubsubskip
\def\beforesubsubspace{\vskip\beforesubsubskip\relax}
\newskip\secskip
\def\secspace{\vskip\secskip\relax}
\newskip\subskip
\def\subspace{\vskip\subskip\relax}
\newskip\insertskip
\def\insertspace{\vskip\insertskip\relax}
\def\sp@ce{\ifx\next*\let\next=\@ssf
               \else\let\next=\@nossf\fi\next}
\def\@ssf#1{\nobreak\secspace\global\ssftrue\nobreak}
\def\@nossf{\nobreak\secspace\nobreak\noindent\ignorespaces}
\def\subsp@ce{\ifx\next*\let\next=\@sssf
               \else\let\next=\@nosssf\fi\next}
\def\@sssf#1{\nobreak\subspace\global\ssftrue\nobreak}
\def\@nosssf{\nobreak\subspace\nobreak\noindent\ignorespaces}
\beforesecskip=24pt plus12pt minus8pt
\beforesubskip=12pt plus6pt minus4pt
\beforesubsubskip=12pt plus6pt minus4pt
\secskip=12pt plus 2pt minus 2pt
\subskip=6pt plus3pt minus2pt
\insertskip=18pt plus6pt minus6pt%
\fontdimen16\tensy=2.7pt
\fontdimen17\tensy=2.7pt
%
% Labels etc for cross referencing macros
%
\def\eqlabel{(\ifappendix\applett
               \ifnumbysec\ifnum\secno>0 \the\secno\fi.\fi
               \else\ifnumbysec\the\secno.\fi\fi\the\countno)}
\def\seclabel{\ifappendix\ifnumapp\else\applett\fi
    \ifnum\secno>0 \the\secno
    \ifnumbysec\ifnum\subno>0.\the\subno\fi\fi\fi
    \else\the\secno\fi\ifnum\subno>0.\the\subno
         \ifnum\subsubno>0.\the\subsubno\fi\fi}
\def\tablabel{\ifappendix\applett\fi\the\tabno}
\def\figlabel{\ifappendix\applett\fi\the\figno}
\def\gac{\global\advance\countno by 1}
%
% Redefinition of footnote macros to lose rule and remove indentation
%

\def\vfootnote#1{\insert\footins\bgroup
\interlinepenalty=\interfootnotelinepenalty
\splittopskip=\ht\strutbox % top baseline for broken footnotes
\splitmaxdepth=\dp\strutbox \floatingpenalty=20000
\leftskip=0pt \rightskip=0pt \spaceskip=0pt \xspaceskip=0pt%
\noindent\smallfonts\rm #1\ \ignorespaces\footstrut\futurelet\next\fo@t}
%
% Redefinition of endinsert to give more controllable
% space around  tables and figures
%
\def\endinsert{\egroup
    \if@mid \dimen@=\ht0 \advance\dimen@ by\dp0
       \advance\dimen@ by12\p@ \advance\dimen@ by\pagetotal
       \ifdim\dimen@>\pagegoal \@midfalse\p@gefalse\fi\fi
    \if@mid \insertspace \box0 \par \ifdim\lastskip<\insertskip
    \removelastskip \penalty-200 \insertspace \fi
    \else\insert\topins{\penalty100
       \splittopskip=0pt \splitmaxdepth=\maxdimen 
       \floatingpenalty=0
       \ifp@ge \dimen@=\dp0
       \vbox to\vsize{\unvbox0 \kern-\dimen@}%
       \else\box0\nobreak\insertspace\fi}\fi\endgroup}   
%
% special macros for display equations
%
% for indentation of turned over lines in mathematics
%
\def\ind{\hbox to \secindent{\hfill}}
%
% for turned over equals sign to left of maths indent
%

%
% for other signs to left of maths indent
%

%
% displayed equation indented 
%
\def\indeqn#1{\alignedfalse\displ@y\halign{\hbox to \displaywidth
    {$\ind\@lign\displaystyle##\hfil$}\crcr #1\crcr}}
%
% displayed equation indented with alignments
%
\def\indalign#1{\alignedtrue\displ@y \tabskip=0pt 
  \halign to\displaywidth{\ind$\@lign\displaystyle{##}$\tabskip=0pt
    &$\@lign\displaystyle{{}##}$\hfill\tabskip=\centering
    &\llap{$\@lign\hbox{\rm##}$}\tabskip=0pt\crcr
    #1\crcr}}
\def\indenteddisplay#1$${\indispl@y{#1 }}
\def\indispl@y#1{\disptest#1\eqalignno\eqalignno\disptest}
\def\disptest#1\eqalignno#2\eqalignno#3\disptest{%
    \ifx#3\eqalignno
    \indalign#2%
    \else\indeqn{#1}\fi$$}
%
% Roman small caps (if in Roman \sc gives small caps)
%

%
% Italic small caps (if in italic \sc gives italic small caps)
%

%
% Bold small caps (if in bold \sc gives bold small caps)
%

%
% Small caps in maths
%

%
% Miscellaneous definitions
%

\def\ns{\noalign{\vskip-3pt}}

%

%
% Bold h bar
%
\def\bhbar{\rlap{\kern1pt\raise.4ex\hbox{\bf\char'40}}\bi{h}}

\def\frac#1#2{{#1\over#2}}
\ifams
\def\lap{\lesssim}
\def\gap{\gtrsim}
\let\le=\leqslant

\let\ge=\geqslant

\else

\def\gap{\;\lower3pt\hbox{$\buildrel > \over \sim$}\;}%
\def\lap{\;\lower3pt\hbox{$\buildrel < \over \sim$}\;}\fi

\chardef\ii="10
\def\tqs{\hbox to 25pt{\hfil}}

\def\Bbbone{1\kern-.22em {\rm l}}
%
% Primes to display summations and products 
% which also have sub or superscripts
%
\def\rp{\raise8pt\hbox{$\scriptstyle\prime$}}
%
% then use \sum^{...}_{...}\rp or \prod^{...}_{...}\rp.
%
% Shadow brackets
%
% Single brackets for normal size only
%

%
% Variable size for display style
%
\def\[#1\]{\setbox0=\hbox{$\dsty#1$}\argwidth=\wd0
    \setbox0=\hbox{$\left[\box0\right]$}\advance\argwidth by -\wd0
    \left[\kern.3\argwidth\box0\kern.3\argwidth\right]}
%
% Variable size for text style
%
\def\lsb#1\rsb{\setbox0=\hbox{$#1$}\argwidth=\wd0
    \setbox0=\hbox{$\left[\box0\right]$}\advance\argwidth by -\wd0
    \left[\kern.3\argwidth\box0\kern.3\argwidth\right]}
%

%
% Square for end of theorems
%

%
\def\pt(#1){({\it #1\/})}
\let\dsty=\displaystyle

%
% Definition for Nuclear Physics Keyword abstract
%
\def\reactions#1{\vskip 12pt plus2pt minus2pt%             
\vbox{\hbox{\kern\secindent\vrule\kern12pt%
\vbox{\kern0.5pt\vbox{\hsize=24pc\parindent=0pt\smallfonts\rm NUCLEAR 
REACTIONS\strut\quad #1\strut}\kern0.5pt}\kern12pt\vrule}}}
%
% Definition for slashed characters
%
\def\slashchar#1{\setbox0=\hbox{$#1$}\dimen0=\wd0%
\setbox1=\hbox{/}\dimen1=\wd1%
\ifdim\dimen0>\dimen1%                        
\rlap{\hbox to \dimen0{\hfil/\hfil}}#1\else                                        
\rlap{\hbox to \dimen1{\hfil$#1$\hfil}}/\fi}
%
% Redefine \textindent for use in \item
%
\def\textindent#1{\noindent\hbox to \parindent{#1\hss}\ignorespaces}
%
% Symbols and curves for use in figure captions
%
\def\opencirc{\raise1pt\hbox{$\scriptstyle{\bigcirc}$}}

\ifams
\def\opensqr{\hbox{$\square$}}

\def\opentridown{\hbox{$\triangledown$}}

\else
\def\opensqr{\vbox{\hrule height.4pt\hbox{\vrule width.4pt height3.5pt
    \kern3.5pt\vrule width.4pt}\hrule height.4pt}}

\def\opentridown{\raise1pt\hbox{$\scriptstyle\bigtriangledown$}}

           %  These produce the
                   %  equivalent open character
           %  to be filled in.
\fi

%
% Redefinition of \cases
%
\def\m@th{\mathsurround=0pt}
%
% Displaystyle now used for first term
%
\def\cases#1{%
\left\{\,\vcenter{\normalbaselines\openup1\jot\m@th%
     \ialign{$\displaystyle##\hfil$&\rm\tqs##\hfil\crcr#1\crcr}}\right.}%
%
% Original version of cases now called \oldcases
%
\def\oldcases#1{\left\{\,\vcenter{\normalbaselines\m@th
    \ialign{$##\hfil$&\rm\quad##\hfil\crcr#1\crcr}}\right.}
%
% Cases with number at end each line (using automatic numbering)
%
\def\numcases#1{\left\{\,\vcenter{\baselineskip=15pt\m@th%
     \ialign{$\displaystyle##\hfil$&\rm\tqs##\hfil
     \crcr#1\crcr}}\right.\hfill
     \vcenter{\baselineskip=15pt\m@th%
     \ialign{\rlap{$\phantom{\displaystyle##\hfil}$}\tabskip=0pt&\en
     \rlap{\phantom{##\hfil}}\crcr#1\crcr}}}
\def\ptnumcases#1{\left\{\,\vcenter{\baselineskip=15pt\m@th%
     \ialign{$\displaystyle##\hfil$&\rm\tqs##\hfil
     \crcr#1\crcr}}\right.\hfill
     \vcenter{\baselineskip=15pt\m@th%
     \ialign{\rlap{$\phantom{\displaystyle##\hfil}$}\tabskip=0pt&\enpt
     \rlap{\phantom{##\hfil}}\crcr#1\crcr}}\global\eqlett=97
     \global\advance\countno by 1}
%
% for equation numbers instead of \eqno
%
\def\eq(#1){\ifaligned\@mp(#1)\else\hfill\llap{{\rm (#1)}}\fi}
\def\ceq(#1){\ns\ns\ifaligned\@mp\fi\eq(#1)\cr\ns\ns}
\def\eqpt(#1#2){\ifaligned\@mp(#1{\it #2\/})
                    \else\hfill\llap{{\rm (#1{\it #2\/})}}\fi}

%
% Automatic numbering of equations
%
\countno=1

\def\aleq{&\rm(\ifappendix\applett
               \ifnumbysec\ifnum\secno>0 \the\secno\fi.\fi
               \else\ifnumbysec\the\secno.\fi\fi\the\countno}
\def\noaleq{\hfill\llap\bgroup\rm(\ifappendix\applett
               \ifnumbysec\ifnum\secno>0 \the\secno\fi.\fi
               \else\ifnumbysec\the\secno.\fi\fi\the\countno}
\def\@mp{&}
\def\en{\ifaligned\aleq)\else\noaleq)\egroup\fi\gac}
\def\cen{\ns\ns\ifaligned\@mp\fi\en\cr\ns\ns}
\def\enpt{\ifaligned\aleq{\it\char\the\eqlett})\else
    \noaleq{\it\char\the\eqlett})\egroup\fi
    \global\advance\eqlett by 1}
\def\endpt{\ifaligned\aleq{\it\char\the\eqlett})\else
    \noaleq{\it\char\the\eqlett})\egroup\fi
    \global\eqlett=97\gac}
%
% abbreviations for Institute of Physics Publishing journals
%
\def\CQG{{\it Class. Quantum Grav.}}

        %1968-87
   %1988 and onwards
     %1968--1988
        %1989 and onwards

           %1975--1988
     %1989 and onwards

                 %1990 and onwards

%
% Other commonly quoted journals
%

\def\NP{{\it Nucl. Phys.}}

\def\PR{{\it Phys. Rev.}}
\def\PRL{{\it Phys. Rev. Lett.}}
\def\PRS{{\it Proc. R. Soc.}}

\def\RMP{{\it Rev. Mod. Phys.}}

\headline={\ifodd\pageno{\ifnum\pageno=\firstpage\hfill
   \else\rrhead\fi}\else\lrhead\fi}
\def\rrhead{\textfonts\hskip\secindent\it
    \shorttitle\hfill\rm\folio}
\def\lrhead{\textfonts\hbox to\secindent{\rm\folio\hss}%
    \it\aunames\hss}
\footline={\ifnum\pageno=\firstpage \hfill\textfonts\rm\folio\fi}
\def\@rticle#1#2{\vglue.5pc
    {\parindent=\secindent \bf #1\par}
     \vskip2.5pc
    {\exhyphenpenalty=10000\hyphenpenalty=10000
     \baselineskip=18pt\raggedright\noindent
     \headfonts\bf#2\par}\futurelet\next\sh@rttitle}%
\def\title#1{\gdef\shorttitle{#1}
    \vglue4pc{\exhyphenpenalty=10000\hyphenpenalty=10000 
    \baselineskip=18pt 
    \raggedright\parindent=0pt
    \headfonts\bf#1\par}\futurelet\next\sh@rttitle} 

\def\article#1#2{\gdef\shorttitle{#2}\@rticle{#1}{#2}} 
\def\review#1{\gdef\shorttitle{#1}%
    \@rticle{REVIEW \ifpbm\else ARTICLE\fi}{#1}}
\def\topical#1{\gdef\shorttitle{#1}%
    \@rticle{TOPICAL REVIEW}{#1}}
\def\comment#1{\gdef\shorttitle{#1}%
    \@rticle{COMMENT}{#1}}
\def\note#1{\gdef\shorttitle{#1}%
    \@rticle{NOTE}{#1}}
\def\prelim#1{\gdef\shorttitle{#1}%
    \@rticle{PRELIMINARY COMMUNICATION}{#1}}
\def\letter#1{\gdef\shorttitle{Letter to the Editor}%
     \gdef\aunames{Letter to the Editor}
     \global\lettertrue\ifnum\jnl=7\global\letterfalse\fi
     \@rticle{LETTER TO THE EDITOR}{#1}}
\def\sh@rttitle{\ifx\next[\let\next=\sh@rt
                \else\let\next=\f@ll\fi\next}
\def\sh@rt[#1]{\gdef\shorttitle{#1}}
\def\f@ll{}
\def\author#1{\ifletter\else\gdef\aunames{#1}\fi\vskip1.5pc
    {\parindent=\secindent  
     \hang\textfonts  
     \ifppt\bf\else\rm\fi#1\par}  
     \ifppt\bigskip\else\smallskip\fi
     \futurelet\next\@unames}
\def\@unames{\ifx\next[\let\next=\short@uthor
                 \else\let\next=\@uthor\fi\next}
\def\short@uthor[#1]{\gdef\aunames{#1}}
\def\@uthor{}
\def\address#1{{\parindent=\secindent
    \exhyphenpenalty=10000\hyphenpenalty=10000
\ifppt\textfonts\else\smallfonts\fi\hang\raggedright\rm#1\par}%
    \ifppt\bigskip\fi}
\def\jl#1{\global\jnl=#1}
\jl{0}%
\def\journal{\ifnum\jnl=1 J. Phys.\ A: Math.\ Gen.\ 
        \else\ifnum\jnl=2 J. Phys.\ B: At.\ Mol.\ Opt.\ Phys.\ 
        \else\ifnum\jnl=3 J. Phys.:\ Condens. Matter\ 
        \else\ifnum\jnl=4 J. Phys.\ G: Nucl.\ Part.\ Phys.\ 
        \else\ifnum\jnl=5 Inverse Problems\ 
        \else\ifnum\jnl=6 Class. Quantum Grav.\ 
        \else\ifnum\jnl=7 Network\ 
        \else\ifnum\jnl=8 Nonlinearity\
        \else\ifnum\jnl=9 Quantum Opt.\
        \else\ifnum\jnl=10 Waves in Random Media\
        \else\ifnum\jnl=11 Pure Appl. Opt.\ 
        \else\ifnum\jnl=12 Phys. Med. Biol.\
        \else\ifnum\jnl=13 Modelling Simulation Mater.\ Sci.\ Eng.\ 
        \else\ifnum\jnl=14 Plasma Phys. Control. Fusion\ 
        \else\ifnum\jnl=15 Physiol. Meas.\ 
        \else\ifnum\jnl=16 Sov.\ Lightwave Commun.\
        \else\ifnum\jnl=17 J. Phys.\ D: Appl.\ Phys.\
        \else\ifnum\jnl=18 Supercond.\ Sci.\ Technol.\
        \else\ifnum\jnl=19 Semicond.\ Sci.\ Technol.\
        \else\ifnum\jnl=20 Nanotechnology\
        \else\ifnum\jnl=21 Meas.\ Sci.\ Technol.\ 
        \else\ifnum\jnl=22 Plasma Sources Sci.\ Technol.\ 
        \else\ifnum\jnl=23 Smart Mater.\ Struct.\ 
        \else\ifnum\jnl=24 J.\ Micromech.\ Microeng.\
   \else Institute of Physics Publishing\ 
   \fi\fi\fi\fi\fi\fi\fi\fi\fi\fi\fi\fi\fi\fi\fi
   \fi\fi\fi\fi\fi\fi\fi\fi\fi}
\def\beginabstract{\insertspace
     \parindent=\secindent\ifppt\textfonts\else\smallfonts\fi
     \hang{\bf Abstract. }\rm }
\def\endabstract{\par
    \parindent=\textind\textfonts\rm
    \ifppt\vfill\fi}

\def\today{\number\day\ \ifcase\month\or
     January\or February\or March\or April\or May\or June\or
     July\or August\or September\or October\or November\or
     December\fi\space \number\year}
\def\date{\ifppt\noindent\textfonts\rm 
     Date: \today\par\goodbreak\bigskip\fi}
%
% Physics Abstracts classification numbers
%

%

%
%%%%%%%%%%%%%%%%%%%%%%%%%%%%%%%%%%%%%%%%%%%%%%%%%%%%%%%%%%%%
%                                                          %
%  Sections, subsections, etc                              %
%                                                          %
%%%%%%%%%%%%%%%%%%%%%%%%%%%%%%%%%%%%%%%%%%%%%%%%%%%%%%%%%%%%
%
\def\section#1{\ifppt\ifnum\secno=0\eject\fi\fi
    \subno=0\subsubno=0\global\advance\secno by 1
    \gdef\labeltype{\seclabel}\ifnumbysec\countno=1\fi
    \goodbreak\beforesecspace\nobreak
    \noindent{\bf \the\secno. #1}\par\futurelet\next\sp@ce}
\def\subsection#1{\subsubno=0\global\advance\subno by 1
     \gdef\labeltype{\seclabel}%
     \ifssf\else\goodbreak\beforesubspace\fi
     \global\ssffalse\nobreak
     \noindent{\it \the\secno.\the\subno. #1\par}%
     \futurelet\next\subsp@ce}
\def\subsubsection#1{\global\advance\subsubno by 1
     \gdef\labeltype{\seclabel}%
     \ifssf\else\goodbreak\beforesubsubspace\fi
     \global\ssffalse\nobreak
     \noindent{\it \the\secno.\the\subno.\the\subsubno. #1}\null. 
     \ignorespaces}
%

%
%%%%%%%%%%%%%%%%%%%%%%%%%%%%%%%%%%%%%%%%%%%%%%%%%%%%%%%%%%%%
%                                                          %
%  Appendices                                              %
%                                                          %
%%%%%%%%%%%%%%%%%%%%%%%%%%%%%%%%%%%%%%%%%%%%%%%%%%%%%%%%%%%%
%
\def\numappendix#1{\ifappendix\ifnumbysec\countno=1\fi\else
    \countno=1\figno=0\tabno=0\fi
    \subno=0\global\advance\appno by 1
    \secno=\appno\gdef\applett{A}\gdef\labeltype{\seclabel}%
    \global\appendixtrue\global\numapptrue
    \goodbreak\beforesecspace\nobreak
    \noindent{\bf Appendix \the\appno. #1\par}%
    \futurelet\next\sp@ce}
\def\numsubappendix#1{\global\advance\subno by 1\subsubno=0
    \gdef\labeltype{\seclabel}%
    \ifssf\else\goodbreak\beforesubspace\fi
    \global\ssffalse\nobreak
    \noindent{\it A\the\appno.\the\subno. #1\par}%
    \futurelet\next\subsp@ce}
\def\@ppendix#1#2#3{\countno=1\subno=0\subsubno=0\secno=0\figno=0\tabno=0
    \gdef\applett{#1}\gdef\labeltype{\seclabel}\global\appendixtrue
    \goodbreak\beforesecspace\nobreak
    \noindent{\bf Appendix#2#3\par}\futurelet\next\sp@ce}
\def\Appendix#1{\@ppendix{A}{. }{#1}}
\def\appendix#1#2{\@ppendix{#1}{ #1. }{#2}}
\def\App#1{\@ppendix{A}{ }{#1}}
\def\app{\@ppendix{A}{}{}}
\def\subappendix#1#2{\global\advance\subno by 1\subsubno=0
    \gdef\labeltype{\seclabel}%
    \ifssf\else\goodbreak\beforesubspace\fi
    \global\ssffalse\nobreak
    \noindent{\it #1\the\subno. #2\par}%
    \nobreak\subspace\noindent\ignorespaces}
%
%%%%%%%%%%%%%%%%%%%%%%%%%%%%%%%%%%%%%%%%%%%%%%%%%%%%%%%%%%%%
%                                                          %
%  Acknowledgments, notes added and foreign abstracts      %
%                                                          %
%%%%%%%%%%%%%%%%%%%%%%%%%%%%%%%%%%%%%%%%%%%%%%%%%%%%%%%%%%%%
%
\def\@ck#1{\ifletter\bigskip\noindent\ignorespaces\else
    \goodbreak\beforesecspace\nobreak
    \noindent{\bf Acknowledgment#1\par}%
    \nobreak\secspace\noindent\ignorespaces\fi}
\def\ack{\@ck{s}}
\def\ackn{\@ck{}}
\def\n@ip#1{\goodbreak\beforesecspace\nobreak
    \noindent\smallfonts{\it #1}. \rm\ignorespaces}
\def\naip{\n@ip{Note added in proof}}
\def\na{\n@ip{Note added}}

%
%  \resume and \zus in Physics in Medicine and Biology only
%

%

%
%%%%%%%%%%%%%%%%%%%%%%%%%%%%%%%%%%%%%%%%%%%%%%%%%%%%%%%%%%%%
%                                                          %
%  Tables                                                  %
%                                                          %
%%%%%%%%%%%%%%%%%%%%%%%%%%%%%%%%%%%%%%%%%%%%%%%%%%%%%%%%%%%
%

%

%

\def\tablecont{\topinsert\global\advance\tabno by -1
    \tablecaption{(continued)}}
\def\tablecaption#1{\gdef\labeltype{\tablabel}\global\widefalse
    \leftskip=\secindent\parindent=0pt
    \global\advance\tabno by 1
    \smallfonts{\bf Table \ifappendix\applett\fi\the\tabno.} \rm #1\par
    \smallskip\futurelet\next\t@b}
\def\t@b{\ifx\next*\let\next=\widet@b
             \else\ifx\next[\let\next=\fullwidet@b
                      \else\let\next=\narrowt@b\fi\fi
             \next}
\def\widet@b#1{\global\widetrue\global\notfulltrue
    \t@bwidth=\hsize\advance\t@bwidth by -\secindent} 
\def\fullwidet@b[#1]{\global\widetrue\global\notfullfalse
    \leftskip=0pt\t@bwidth=\hsize}                  
\def\narrowt@b{\global\notfulltrue}
\def\align{\catcode`?=13\ifnotfull\moveright\secindent\fi
    \vbox\bgroup\halign\ifwide to \t@bwidth\fi
    \bgroup\strut\tabskip=1.2pc plus1pc minus.5pc}
\def\endalign{\egroup\egroup\catcode`?=12}

%
% Use \L{#}, \R{#} and \C{#} to specify left, right or centred
% columns immediately after \table. For example
% \align\L{#}&&\L{#}\cr gives the preamble for a table with
% all columns aligned left, \align\L{#}&\C{#}&\R{#}\cr
% gives a table with 3 columns, the first aligned left, the second
% centred and the third aligned right.
%

%
%  Rules for tables \br at top and bottom
%  \mr to separate headings from entries
%

%

%
% Definitions for centring headings over several columns
% \centre{4}{Results for helium} will centre
% Results for helium over four columns
% \crule{4} will produce a rule centred over four columns
% to go below a centred heading
%

%

\catcode`?=13
\def\lineup{\setbox0=\hbox{\smallfonts\rm 0}%
    \digitwidth=\wd0%
    \def?{\kern\digitwidth}%
    \def\\{\hbox{$\phantom{-}$}}%
    \def\-{\llap{$-$}}}
\catcode`?=12
%
% Macros for two parts of a table of equal width side by side
% \table{caption}[w]
% \sidetable{first part}{second part}
% \endtable
% Use \table preamble for tables of 31picas width
%
\def\sidetable#1#2{\hbox{\ifppt\hsize=18pc\t@bwidth=18pc
                          \else\hsize=15pc\t@bwidth=15pc\fi
    \parindent=0pt\vtop{\null #1\par}%
    \ifppt\hskip1.2pc\else\hskip1pc\fi
    \vtop{\null #2\par}}} 
\def\lstable#1#2{\everypar{}\tempval=\hsize\hsize=\vsize
    \vsize=\tempval\hoffset=-3pc
    \global\tabno=#1\gdef\labeltype{\tablabel}%
    \noindent\smallfonts{\bf Table \ifappendix\applett\fi
    \the\tabno.} \rm #2\par
    \smallskip\futurelet\next\t@b}
\def\inctabno{\global\advance\tabno by 1}
%
%%%%%%%%%%%%%%%%%%%%%%%%%%%%%%%%%%%%%%%%%%%%%%%%%%%%%%%%%%%%
%                                                          %
%  Figures                                                 %
%                                                          %
%%%%%%%%%%%%%%%%%%%%%%%%%%%%%%%%%%%%%%%%%%%%%%%%%%%%%%%%%%%%
%

%

%
\def\figure#1{\figc@ption{#1}\bigskip}
\def\figc@ption#1{\global\advance\figno by 1\gdef\labeltype{\figlabel}%
   {\parindent=\secindent\smallfonts\hang
    {\bf Figure \ifappendix\applett\fi\the\figno.} \rm #1\par}}
%
%%%%%%%%%%%%%%%%%%%%%%%%%%%%%%%%%%%%%%%%%%%%%%%%%%%%%%%%%%%%
%                                                          %
%  Reference lists                                         %
%                                                          %
%%%%%%%%%%%%%%%%%%%%%%%%%%%%%%%%%%%%%%%%%%%%%%%%%%%%%%%%%%%%
%
\def\refHEAD{\goodbreak\beforesecspace
     \noindent\textfonts{\bf References}\par
     \let\ref=\rf
     \nobreak\smallfonts\rm}
\def\numreferences{\refHEAD\parindent=30pt
     \everypar{\hang\noindent\frenchspacing\rm}
     \secspace}
\def\rf#1{\par\noindent\hbox to 21pt{\hss #1\quad}\ignorespaces}
%

%

%
% reference to a journal article in numerical system
%
\def\numrefjl#1#2#3#4#5{\par\rf{#1}#2 {\it #3 \bf #4} #5\par}
%
% reference to a book or report in numerical system
%
\def\numrefbk#1#2#3#4{\par\rf{#1}#2 {\it #3} #4\par}
%
%%%%%%%%%%%%%%%%%%%%%%%%%%%%%%%%%%%%%%%%%%%%%%%%%%%%%%%%%%%%
%                                                          %
%  Theorems, lemmas, etc                                   %
%                                                          %
%%%%%%%%%%%%%%%%%%%%%%%%%%%%%%%%%%%%%%%%%%%%%%%%%%%%%%%%%%%%
%

%
% NB \note#1 is used to give a Note (as opposed to a paper or letter)
% in PMB therefore use commands \notes#1 for numbered Note
% instead of \note 
%

%
\catcode`\@=12
%
% Parameter values for `Preprint' style 
%
\def\pptstyle{\ppttrue\headsize{17}{24}%
\textsize{12}{16}%
\smallsize{10}{12}%
\hsize=37.2pc\vsize=56pc
\textind=20pt\secindent=6pc}
%
% Parameter values for `Journal' style 
%

%
% Parameter values for `Eleven point' style 
%

%
% Parameter values for `Large size' style 
%

%
\parindent=\textind

\pptstyle
\jl{6}

\font\small=cmr10
\font\teni=cmmi10
\font\sevenrm=cmr7
\font\tenex=cmex10

\def\oneskip{\vskip\baselineskip\noindent}

\input psfig.tex

\topical{Astrophysical evidence for the existence of black holes}

\author{A Celotti, J C Miller and D W Sciama}

\address{SISSA, Via Beirut 2-4, 34013 Trieste, Italy.}

\beginabstract Following a short account of the history of the idea of
black holes, we present a review of the current status of the search for
observational evidence of their existence aimed at an audience of
relativists rather than astronomers or astrophysicists. We focus on two
different regimes: that of stellar-mass black holes and that of black
holes with the masses of galactic nuclei.
\endabstract

%\submitted

\date

\section{Introduction}

In his famous article of 1784, which is seen as being the beginning of the
story of black holes, John Michell [1] wrote:
\bigskip

{\narrower\narrower{\noindent
If there should really exist in nature any [such] bodies, $\ldots$ we
could have no information from sight; yet, if any other luminous bodies
should happen to revolve about them we might still perhaps from the
motions of these revolving bodies infer the existence of the central ones
with some degree of probability, as this might afford a clue to some of
the apparent irregularities of the revolving bodies, which would not be
easily explicable on any other hypothesis.}

} 
\bigskip 
\noindent 
 There at the very beginning, the theoretically-predicted properties of
(Newtonian) black holes were discussed together with a carefully-worded
statement about how it might be determined observationally whether such
objects do {\it in fact} exist. Following Michell's paper and the
subsequent repetition of his arguments by Laplace [2] in his book of 1796,
there is a long gap until the present century when, with the coming of
general relativity, the theoretical discussion of black holes started
anew. The observational search had to wait rather longer, until the
development of radio astronomy (from the late 1940s onwards) and X-ray
astronomy (from the 1960s onwards).

The basic theory of black holes is now well-understood, but the path to
this understanding has been long and tortuous. It has been beautifully
described by Israel [3] in the book ``300 Years of Gravitation''. The
first black-hole solution of Einstein's field equations was discovered in
heroic circumstances by Karl Schwarzschild [4] who was looking for the
exact solution for a point mass in otherwise empty space. He discovered
his solution only two months after the publication in 1915 of Einstein's
definitive paper on general relativity [5], while serving in the German
army. He died shortly thereafter.

The black hole property of the Schwarzschild solution was soon recognised;
it was certainly known to Sir Oliver Lodge in 1921 [6]. However, progress
in understanding (and accepting) black holes was slow - people were
unwilling to accept that actual physical objects would ever collapse to
such an extreme state. Another stumbling block rested on a mathematical
misunderstanding springing from the obvious fact that in the usually
adopted Schwarzschild co-ordinates the metric becomes singular at the
event horizon $r = r_s = 2GM/c^2$ (where $M$ is the mass of the black
hole). Also, a further cause for the delay in understanding came from the
fact that in the thirties people did not seem to pay much attention to the
discoveries reported by other people.

One discovery which was attended to, but was found to be rather alarming,
was Chandrasekhar's crucial result of 1931 [7] that relativistic
degeneracy of electrons could not prevent a cold body more massive than
$\sim 1.44\,M_{\odot}$ (where $M_{\odot}$ denotes the mass of the sun) from
collapsing indefinitely. Much later, after the alarm was dissipated,
Chandrasekhar was awarded a share in the Nobel prize for this discovery. An
influential figure whose alarm at the time was acute was Sir Arthur
Eddington, who famously said in 1935 [8] after referring to Chandrasekhar's
result: ``Various accidents may intervene to save the star, but I want more
protection than that. I think there should be a law of Nature to prevent a
star from behaving in this absurd way!''.

In 1933 Lemaitre [9] had already noted that the Schwarzschild singularity
was a mathematical artefact on the same footing as certain singularities
in cosmological metrics which puzzled people for many years. Yet we find
six years later no less a person than Einstein himself [10] writing a
fifteen page paper, full of calculations arguing that a physical system
would necessarily have to be larger than the radius at which the
``undesirable'' Schwarzschild singularity would otherwise occur.

In that same year (1939) Oppenheimer and Snyder [11] wrote a definitive
paper in a prominent journal (Physical Review) showing that a spherically
symmetric pressure-free mass would collapse indefinitely, and correctly
described the effect on light propagation in the neighbourhood of the
event horizon. Yet the authoritative textbook on general relativity by
Bergmann [12] published three years later does not mention the crucial
Oppenheimer-Snyder result, but does spend a whole page on Einstein's
paper.

The modern epoch of understanding commenced in the nineteen fifties,
partly from the introduction of coordinate systems in which the metric is
regular everywhere except at the origin (of which the Kruskal metric [13]
is the best known) and partly from John Wheeler's relentless emphasis on
the importance of understanding the final stages of gravitational collapse
[14]. Other highlights were Kerr's discovery [15] of an exact solution for
rotating black holes, and Israel's uniqueness theorem [16] for
non-rotating black holes, soon to be extended by Carter and others [17] to
the rotating case. Another spectacular result was Penrose's 1965
singularity theorem [18] which showed that formation of physical
singularities (such as that at $r = 0$ for Schwarzschild black holes) was
a generic feature of continued collapse, contradicting the widely held
view that the occurrence of a singularity was a special feature related to
exact spherical symmetry. Penrose's result was soon extended to
cosmological space-times by Hawking [19] and by Geroch [20].

With hindsight we can see that, once it was understood what is the
appropriate level of mathematical sophistication to adopt for research on
black holes, the stage was set for the great complex of discoveries that
were made in the nineteen fifties and sixties, mainly by the considerable
number of highly talented young mathematicians and physicists who were
attracted to enter the field in those years.

Nevertheless there was still a shock to come. The widening of people's
mental horizons did not prepare them for perhaps the greatest discovery
yet made about black holes, namely the Hawking radiation [21] which they
are predicted to emit when quantum effects are taken into account and
which permits a black hole thermodynamics to be set up. This wonderful
result, published in 1974, was received with widespread incredulity, but
only for a short time. Perhaps people were learning the important lesson
that they should consider what a well-established theory says in an
open-minded way and not be blinded by possibly mistaken personal
intuition.

So, where do we stand on the question of whether black holes as described
by the mathematical solutions discussed above actually exist in nature?
For the remainder of this article, we will be concentrating on this
question.

In the second half of the twentieth century, relativistic astrophysics in
general and the study of black holes in particular has changed from being
a fairly marginal minority interest as far as astronomers were concerned,
to being in the central mainstream of astrophysical research. This change
has been closely connected with advances in technology: the advent of
radio and X-ray astronomy (as already noted) and also,later, the rise of
numerical computing which has allowed relativistic calculations to be made
for more complex physical situations than would be possible analytically,
permitting closer confrontation with observations.

What are the circumstances in which astrophysical black holes are thought
likely to be formed?  The ``mean density'' $\bar\rho$ of a black hole (its
mass $M$ divided by ${4 \over 3}\pi r_s^3$) is proportional to $1/M^2$.
For a $1\,M_{\odot}$ black hole, $\bar\rho \sim 10^{16}\,$g$\,$cm$^{-3}$,
forty times nuclear matter density, whereas for a black hole of
$10^8\,M_\odot$, $\bar\rho \sim 1\,$g$\,$cm$^{-3}$, the density of water.
The conditions required for matter to form a small black hole are much
more extreme than for a large one. Three main regimes are being discussed:

\medskip

\item{(a)} Stellar-mass black holes formed after the death of some normal
stars.

\medskip

\item{(b)} Super-massive black holes ($\sim 10^6 - 10^{10}\,M_\odot$)
formed in the centres of galaxies as a result of the processes of galactic
dynamics. (Collapse of super-massive stars or relativistic star clusters
might also produce high-mass black holes.)

\medskip

\item{(c)} Black holes formed as a result of fluctuations or phase
transitions in the early universe when conditions were so extreme that
black holes of all masses might have been produced.

\medskip
\noindent
 We will be concentrating here on the stellar mass black holes of class
(a) and the galactic-centre black holes of class (b), since these are the
ones most connected with observational searches. However, primordial black
holes of class (c) have periodically attracted considerable attention
(most recently in connection with gravitational microlensing [22]). For a
review of primordial black holes, see the article by Carr [23].

As noted by Blandford in his 1987 review article on astrophysical black
holes [24] (to which we refer the reader for coverage of many details
which we will not be dealing with here), the discussion of whether or not
a particular object is indeed a black hole has a rather different
character in the case of the stellar-mass black hole candidates from that
for the super-massive ones since, for stellar-mass candidates, the main
alternatives to black holes are neutron stars which are rather
clearly-specified objects about which much is known (although see our
discussion in Section 2), whereas for the super-massive candidates the
alternatives (dense star clusters, superstars, ``magnetoids'', etc.) have
less clearly constrained properties. 

In this article, we make no attempt at any complete coverage of the very
extensive literature involved; our aim is to give a general outline of the
main ideas and lines of investigation in a way particularly aimed at
relativists who are not specialist astronomers. The references have been
limited to a minimum and are focussed on mentioning main key papers
supplemented by a few quite complete but more specific reviews and some
recent publications, which do include more extensive references. We
apologise in advance to all of those colleagues whose work is not directly
referenced.

\section{Stellar-mass black holes}

The study of stellar-mass black hole candidates has been intimately
connected with the development of X-ray astronomy and so we will first
give a brief history of this. 

\subsection{Early history}

Since the Earth's atmosphere is very effective in shielding us from
X-rays, it is necessary to put detectors on board space-craft or
high-altitude balloons in order to make X-ray observations of astronomical
objects. The first such observations were made in 1962 by Giacconi,
Gursky, Paolini and Rossi [25] using detectors on board an Aerobee
space-rocket and they discovered the existence of discrete sources of
X-rays outside the solar system. A succession of subsequent rocket and
balloon experiments then confirmed the earlier results and obtained more
data. By the end of the 1960s around twenty sources had been identified
including Cygnus X-1 which is particularly bright and was found to vary
with time.

In 1966 an old 12th-13th magnitude star was identified as the optical
counterpart of the brightest X-ray source Scorpio X-1. It was not
plausible that the X-rays could be coming from the optical star itself
and, in the following year (1967, the year in which pulsars were first
discovered), a model was proposed [26] in which they were explained as
originating from gas which becomes very hot in the process of being
accreted from the observed star onto an undetected binary companion, a
compact object which was taken to be a neutron star. To be precise, by
``compactness'' we mean the ratio of the actual radius of the object to
its Schwarzschild radius $r_s$: the smaller this ratio is, the more
compact the object is said to be. The term ``compact object'' is used to
refer to white dwarfs, neutron stars and black holes, but we should
emphasize that while white dwarfs are much more compact than ordinary
stars like the sun, they are not nearly as compact as neutron stars and
black holes. Some relative sizes are shown in Fig.~1: neutron stars can
come rather close to being black holes.
{ \midinsert
\null\oneskip
\centerline{\psfig{figure=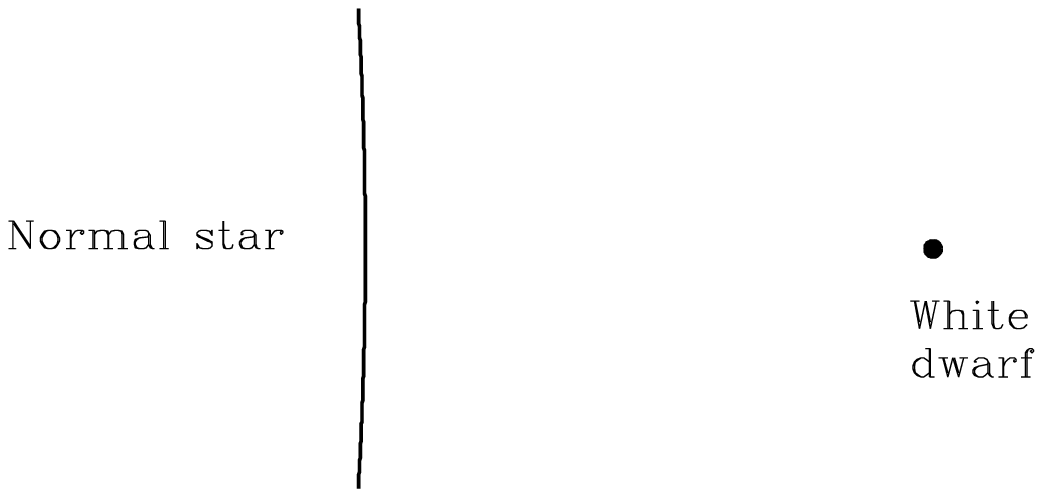,width=14truecm}\hskip 2.7truecm}
\vskip -0.45cm\noindent
\hskip 1.5truecm \vbox{\hrule width 12truecm height 1pt}

\centerline{\psfig{figure=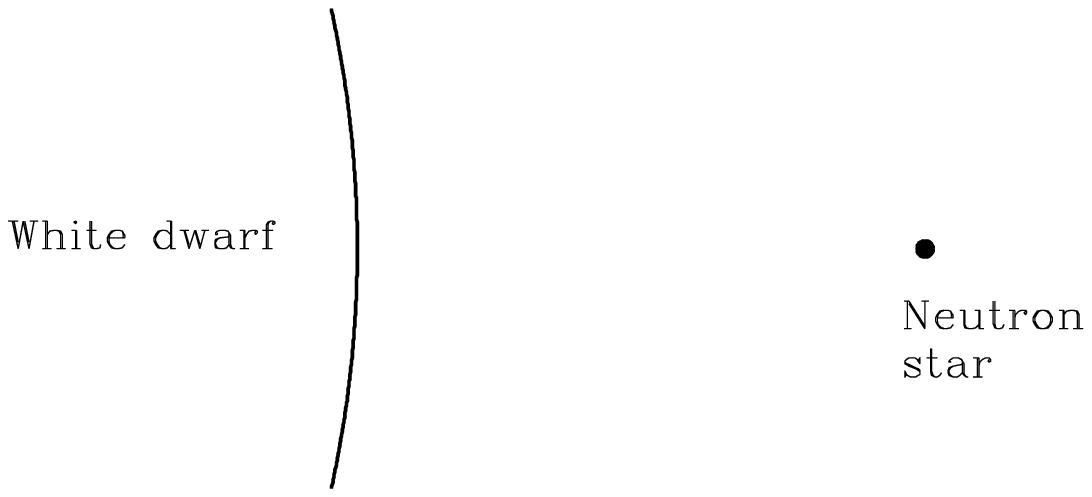,width=14truecm}\hskip 2.7truecm}
\vskip -0.45cm\noindent
\hskip 1.5truecm \vbox{\hrule width 12truecm height 1pt}

\centerline{\psfig{figure=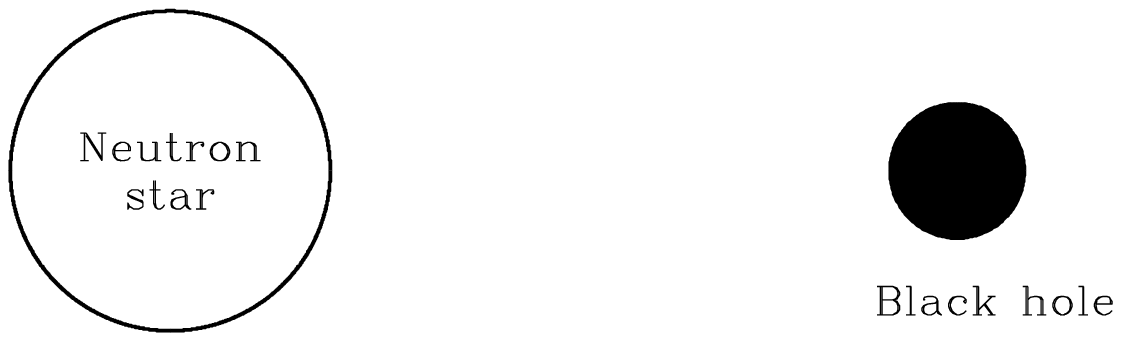,width=14truecm}\hskip 2.7truecm}
\medskip
\noindent
{\baselineskip=12pt
\narrower\narrower\noindent
\textfont0=\small \textfont1=\teni \textfont3=\tenex
\scriptfont0=\sevenrm
\small
Fig.~1. Relative sizes of normal stars, white dwarfs, neutron
stars and black holes having similar masses (we have taken
1.4$\,M_\odot$). Note that while white dwarfs are much more compact
than normal stars, they are not nearly as compact as neutron stars
or black holes which, however, come rather close together.

}
\endinsert
}

In 1968, Prendergast and Burbidge [27] argued that the accreted material
in this model would be carrying so much orbital angular momentum
that it would not fall onto the compact object along radial paths but
rather would be drawn into a thin accretion disc around the compact object
(see Fig.~2) and then slowly spiral inwards as a result of viscous drag.
The viscosity also heats the disc up to temperatures at which it emits the
observed radiation. (Note that this picture can apply only if the magnetic
field of the compact object is not too high; a strong dipole magnetic
field would cause the flow to be funnelled down the magnetic axis.) The
accretion may be initiated by the optical star swelling up to fill an
equipotential surface called the Roche lobe and then losing matter to its
companion via the crossing-point in this surface. Gravitational potential
energy liberated as the accreted matter spirals down the gravitational
potential well of the compact object is converted partly into kinetic
energy (mainly rotational) and partly into thermal energy, some of which
is then radiated. This can be an extremely efficient energy generation
mechanism, particularly if the compact object is a black hole.

{
\midinsert
\null\oneskip
\noindent
\centerline{\psfig{figure=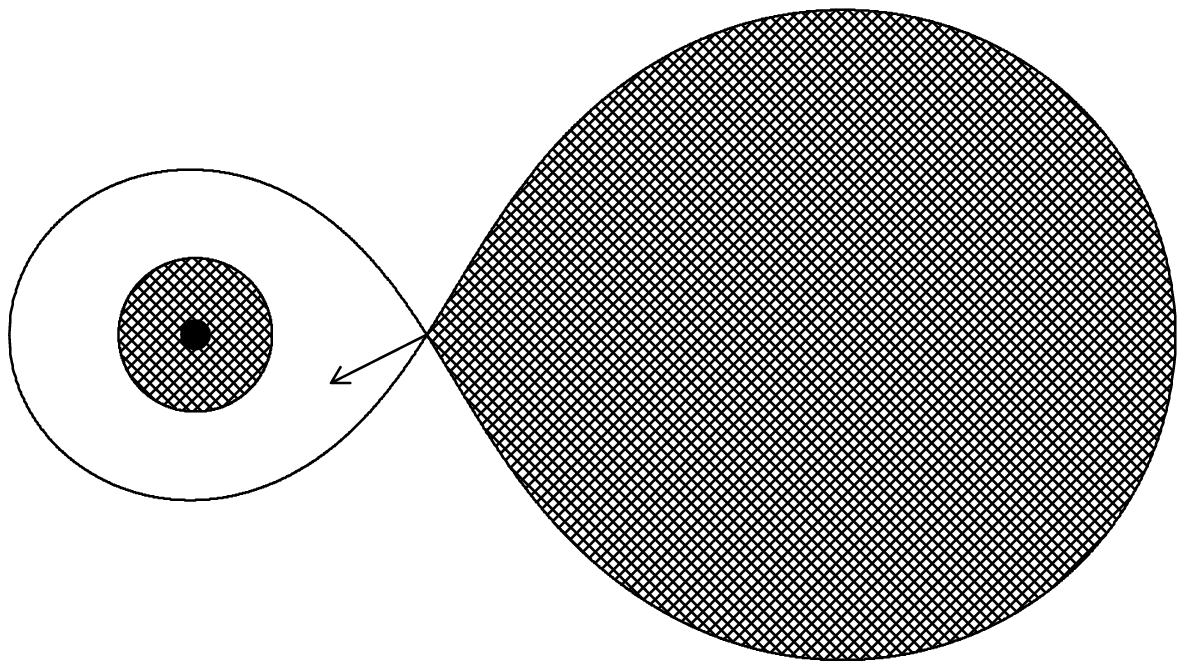,width=9.2truecm}
\hskip 1 truecm}
\oneskip
\narrower\narrower\noindent
{\baselineskip=12pt
\small
Fig.~2. View down the rotation axis of a binary system in which 
accretion is taking place. The secondary star has swollen up to fill an 
equipotential surface called the Roche lobe and is spilling matter across 
onto the compact object. Because of its high angular momentum, this 
accreted material forms itself into a rotating disc.
\medskip
}
\endinsert
}

However, up to this point, there had not yet been conclusive observational
evidence that any observed X-ray sources were, in fact, in close binary
systems or were associated with compact objects. The evidence that this
was so came from the Uhuru satellite launched in 1970. Uhuru was entirely
devoted to X-ray observations and continued to operate for more than three
years during which time it found more than three hundred sources and
established that many of them were indeed associated with binary systems
consisting of an optically-observed star together with an
optically-invisible companion [28]. Also, short timescale variability was
seen in the X-ray emission of many sources which, assuming that variations
cannot occur on a timescale shorter than the light crossing time,
indicated that they did contain compact objects. 

\subsection{Cygnus X-1}

During the 1970s and 1980s, particular attention was focussed on the
source Cygnus X-1, which appeared to be the strongest candidate for
containing a black hole. This source was positively identified with a
binary system consisting of a high-mass OB supergiant star, HDE 226868,
orbiting an unseen companion with an orbital period of 5.6 days. It
showed X-ray variability on a range of timescales extending down to one
millisecond, indicating that the companion is extremely compact and must
be either a neutron star or a black hole [29].  How can one distinguish
between these two possibilities? Neutron stars cannot have arbitrarily
large mass; there is a maximum above which the pressure can no longer
balance gravity. This maximum is currently thought to be somewhere between
$1.4\,M_{\odot}$ and $2.5\,M_{\odot}$ if the neutron star is non-rotating
(or rotating only very slowly) and may be raised by up to $25\%$ if it is
rotating rapidly. If one could determine that the mass of a very compact
object is above the maximum for a neutron star, then it would presumably
have to be a black hole. This is the line of reasoning that was followed
with Cygnus X-1 and with various subsequent black hole candidates.

How can the mass of the compact object be established from observations? 
We know that HDE 226868 is a member of a binary system because its
spectrum shows systematic Doppler shifts which are consistent with it
moving on a binary orbit under the influence of the unseen companion.
From the Doppler shift data, a radial velocity curve can be constructed,
giving the variation with time of the component of the star's velocity
along the line of sight. From this one can extract the orbital period
$P$, the semi-amplitude of the curve $K$ and, in principle, the
eccentricity of the orbit (although this is extremely small in the case of
Cygnus X-1 and, for simplicity, we will neglect it in the following
discussion). Next, we make use of Kepler's laws of orbital motion applied
to the layout shown schematically in Fig.~3. 
{
\midinsert
\noindent
\centerline{\psfig{figure=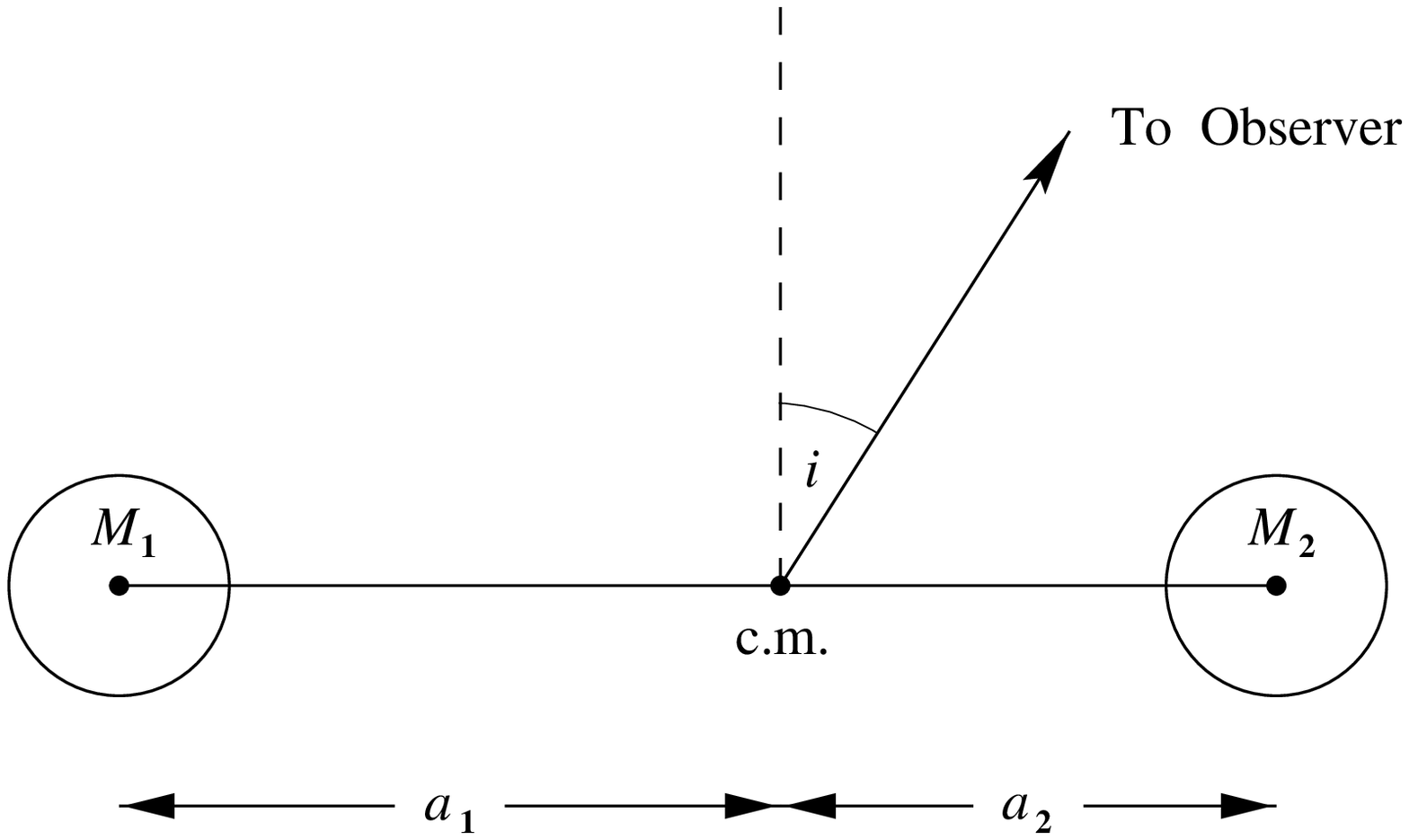,width=9.2truecm,angle=270}
\hskip 1 truecm} 
\oneskip
\oneskip
\narrower\narrower\noindent
{\baselineskip=12pt 
\textfont1=\teni \scriptfont0=\sevenrm
\small
Fig.~3. Schematic picture of the binary parameters as viewed in the
orbital plane of the system. The compact object has mass $M_1$, the
optical star has mass $M_2$ and distances are measured from the centre of
mass (c.m.) of the system. 

\medskip
} 
\endinsert 
} 
Using Kepler's second law, we have that $K = 2\pi a_2 \sin i / P$ and
then, using Kepler's third law, one can obtain a quantity called the {\it
mass function}: 
$$ 
f(M_1) = { {{\left( M_1 \sin i \right)}^3} \over {{\left( M_1 + M_2
\right)}^2} } 
= { {PK^3} \over {2\pi G} }.
$$
 For Cygnus X-1, $f(M_1) = (0.252 \pm 0.010)\,M_{\odot}$. In order to
determine the mass of the compact object, it is necessary to supply two
further pieces of information: the mass of the optical star and the angle
of inclination of the orbit (or other information which allows these to be
calculated).  Typically, OB supergiants have masses greater than
$20\,M_{\odot}$ and taking plausible values for the mass of HDE 226868 and
for $\sin i$ led to the conclusion that the mass of the compact object is
probably in the range $9 - 15\,M_{\odot}$, which is well above the maximum
for neutron stars quoted earlier. It is very likely that Cygnus X-1 {\it
does} contain a black hole with a mass of more than $9\,M_{\odot}$ but one
ought to be cautious and ask what is the minimum possible value of $M_1$
consistent with the data.

While OB supergiants do typically have masses greater than
$20\,M_{\odot}$, it is also possible for them to be considerably less
massive without there being any significant change in the spectrum.
However, there is another constraint for HDE 226868 which comes from
requiring that it should be able to produce the observed luminosity. Its
distance from us can be calculated by considering interstellar absorption
observed in the optical spectrum and this gave around 2.5 kiloparsecs (1
parsec $\simeq 3 \times 10^{18}$ cm) with a firm lower limit of about 2
kiloparsecs. One then asks what would be the minimum mass for the star in
order that it could produce the observed luminosity if it is 2 kiloparsecs
away. According to stellar structure calculations of the time, this
came out at $8.5\,M_{\odot}$. A lower limit for the mass of the compact
object can then be obtained by noting that while the angle of inclination
of the orbit is not known, $\sin i$ must certainly be $\le 1$. This then
gives $M_1 \gap 3.3\,M_{\odot}$ which is above the presumed maximum for a
non-rotating neutron star making Cygnus X-1 a good candidate for
containing a black hole. 

In fact, it is possible to do rather better than this if one has
observational data of sufficient quality and a detailed analysis of the
Cygnus X-1 system was presented in 1986 by Gies and Bolton [30], combining
together data from fifty-five observations spread over ten years. By
measuring the rotation-broadening of the optical star's absorption
spectrum and making certain assumptions about the system, it is possible
to derive a value for the {\it mass ratio} $q = M_1/M_2$. Modelling the
ellipsoidal modulation of the light curve (resulting from the fact that a
non-spherical emitter presents a time-varying effective emitting area to
the observer as it goes round its orbit) then allows $\sin i$ to be
calculated and hence all of the parameters can be determined consistently
with the assumptions made. Proceeding in this way, Gies and Bolton derived
a lower limit for $M_1$ of $7\,M_{\odot}$ with a preferred value of
$16\,M_{\odot}$.

However, some doubts remain about Cygnus X-1 which make it difficult to
determine an accurate mass for the compact object. The optical star is
highly evolved and of uncertain mass, does not seem to be filling its
Roche lobe and may not be rotating synchronously with the orbital motion;
there is no velocity information about the compact object itself. (In the
early days there was also a suggestion that there might have been a third
object in the system [31].) Now, there is a new generation of rather
different candidates for which the determination of accurate masses is
easier. 

\subsection{Soft X-ray transients}

The uncertainties with Cygnus X-1 had mainly been related to the fact that
it is in a high-mass X-ray binary ({\it i.e.} the ``normal'' component is
a high-mass star). At first sight, low-mass X-ray binaries (LMXBs) would
seem to be more straightforward to deal with but they suffer from the
major difficulty that the optical light is dominated by that coming from
the X-ray irradiated accretion disc which is much brighter than the
optical companion, normally making the latter impossible to study. This is
not a problem, however, if the source is transient, emitting X-rays for
only part of the time. The best current stellar-mass black hole candidates
are {\it soft X-ray transients} (SXTs), a sub-class of LMXBs which are
X-ray luminous for around six months in every 10 -- 50 years [32]. During
quiescence, the accretion disc becomes extremely faint and it is then
possible to carry out detailed photometry and spectroscopy of the optical
companion. This is still not an easy task, however, because these objects
are themselves very faint and it is only recently that the equipment has
existed on large telescopes to enable such studies to be made. It is
widely considered that the current strongest candidate is the source V404
Cygni which is associated with a binary system having a 6.5 day orbital
period and has an interval between outbursts of 25 years (or it may be one
half of that). The last outburst was in 1989.

For this sort of system, the assumptions of corotation and of the optical
star filling its Roche lobe seem to be rather good and by combining the
measurements of rotation-broadening of the absorption spectrum and the
ellipsoidal variation of the light curve (best done here at infra-red
wavelengths in order to reduce contamination from the accretion disc) it
is possible to make a reliable determination of all of the binary system
parameters. Proceeding in this way, Shahbaz {\it et al.} [33] determined
the masses of both components in several SXTs. For V404 Cygni, they
measured the mass of the compact object as $12 \pm 2 \, M_\odot$ which is
well above the usually accepted mass limit for a neutron star. In the next
section, we say something more about this mass limit.

\subsection{The maximum mass for neutron stars}

How much should one trust the upper mass limit for neutron stars? This has
been obtained using equations of state for material at very high densities
extending above that of nuclear matter where the physics is not very well
understood and is poorly constrained by experimental data. Different
equations of state give a considerable range of values for the neutron
star maximum mass.

In 1974 Rhoades and Ruffini [34] presented an argument designed to
circumvent the uncertainty in the equation of state of matter at very high
densities and to derive a firm upper mass limit for neutron stars, on the
basis only of knowledge which could be considered as really secure. They
noted that the equation of state {\it is} thought to be quite well-known
at lower densities and so it could be taken as fixed for densities $\rho$
less than some fiducial value $\rho_0$, while for $\rho > \rho_0$ it was
varied so as to obtain the maximum possible mass consistent with the
following assumptions: 
\medskip
\item{(i)} Gravity is well described by general relativity;
\medskip
\item{(ii)} The equation of state depends on only one parameter (the
pressure $p = p(\rho)$);
\medskip
\item{(iii)} The matter is microscopically stable: $dp/d\rho \ge 0$;
\medskip
\item{(iv)} The equation of state satisfies the causality condition that
the speed of sound $c_s$ should be less than the speed of light. This was
expressed as $c_s^2 = dp/d\rho \le c^2$. 

\medskip\noindent
Doing this and taking $\rho_0 = 4.6 \times 10^{14}\,{\rm
g\,cm}^{-3}$, they found the maximum allowed mass to be $M_{max}
\simeq 3.2 M_\odot$. For other values of $\rho_0$, the result
scales roughly as
$$
M_{max} \simeq 3.2 \left( { { \rho_0 } \over 
{ 4.6 \times 10^{14}\,{\rm g\,cm}^{-3} } } \right)^{\! -1/2} M_\odot.
$$
%\eject
\noindent
A number of questions now arise:
\medskip
\item{(a)} Is the causality constraint (iv) reasonable?  As pointed out
by Hartle [35], $(dp/d\rho)^{1/2}$ is the {\it phase velocity} for small
disturbances and is equal to the {\it group velocity} (which is the
relevant quantity for causality considerations) only if the medium is {\it
non-dispersive} which is not the case for neutron-star matter. However,
even without condition (iv), it is still possible to obtain an upper
limit, and this is given by
$$
M_{max} \simeq 5.3 \left( { { \rho_0 } \over
{ 4.6 \times 10^{14}\,{\rm g\,cm}^{-3} } } \right)^{\! -1/2} M_\odot.
$$
\medskip
\item{(b)} Do we trust out knowledge of the equation of state up to $4.6
\times 10^{14}\,{\rm g\,cm}^{-3}$ for {\it bulk} high-density matter? A
lower threshold figure would be safer. With $\rho_0$ referred to
$10^{14}\,{\rm g\,cm}^{-3}$, the basic formula becomes
$$
M_{max} \simeq 6.8 \left( { { \rho_0 } \over
{ 10^{14}\,{\rm g\,cm}^{-3} } } \right)^{\! -1/2} M_\odot,
$$
or
$$ 
\hbox{ \ \ \ \ \ \ \ } \simeq 11.4 \left( { { \rho_0 } \over
{ 10^{14}\,{\rm g\,cm}^{-3} } } \right)^{\! -1/2} M_\odot,
$$
when combined with (a) above.

\medskip
\item{(c)} Then there is rotation. Friedman and Ipser [36] showed that if
one allows for uniform rotation up to the shedding limit
$$
M_{max} \simeq 8.4 \left( { { \rho_0 } \over
{ 10^{14}\,{\rm g\,cm}^{-3} } } \right)^{\! -1/2} M_\odot,
$$
with condition (iv) or
$$
\hbox{ \ \ \ \ \ \ \ } \simeq 14 \left( { { \rho_0 } \over
{ 10^{14}\,{\rm g\,cm}^{-3} } } \right)^{\! -1/2} M_\odot,
$$
without condition (iv). 

\medskip\noindent 
At this point, it is no longer clear that the situation is safe even for
V404 Cygni. Standard realistic neutron star equations of state do, in
practice, give masses roughly satisfying the original condition $M_{max}
\lap 3.2\,M_\odot$ even with rotation but now we are back to asking how
completely we trust the picture which they represent. Are we really sure
that compact stars at these densities do consist of a mixture of neutrons,
protons, electrons, mesons, hyperons and other particles held together
just by gravity or could another picture be correct?  (While it is the
strong force, and not gravity, which holds together atomic nuclei,
standard neutron star models do not envisage the strong force playing any
confining role on a macroscopic scale. In this sense, they are very
different from ``giant nuclei''.)

A radically different viewpoint was presented by Witten in 1984 [37] with
the introduction of the {\it strange star model}. Quantum chromodynamics
contains the idea of {\it confinement} by the strong force which is
normally thought of in terms of quarks being confined within nucleons. The
idea of the strange star model rests on the hypothesis that strange quark
matter (composed of almost equal numbers of up, down and strange quarks)
might be the absolute ground state of baryonic matter even at zero
pressure. Strange stars would be essentially giant nucleons with quarks
being confined inside them but able to move freely in the false vacuum
which extends throughout the interior. They could have masses and radii
similar to those of standard neutron stars and have been advocated as a
viable alternative model for pulsars, although there may be a problem over
explaining glitches.

Strange stars are ``safe'' as far as the mass limit is concerned. Although
their equation of state is very different from that for standard neutron
star matter, the maximum mass is within the standard range (it is $\sim
2.0\,M_\odot$ for non-rotating models with some variation depending on
uncertain parameter values). However, some effective field theories of the
strong force allow for it not only to confine quarks in the normal way but
also to confine {\it nucleons} (neutrons and protons) at densities well
below that of nuclear matter thus giving an equation of state different
from the standard one at densities below the values normally taken for
$\rho_0$. Models based on this idea were introduced in 1989 by Bahcall,
Lynn \& Selipsky [38] who named them {\it Q-stars} (although note that the
``Q'' here does not stand for ``quark'' but for a conserved particle
number). These are {\it not} safe for the mass limit and might, in
principle, have {\it very} high masses up to more than $100\,M_\odot$ as
well as being extremely compact with radii down to only 1.4 times those of
equivalent black holes [39]. For low mass models, the Q-stars would have
an almost constant density throughout but for higher masses the density
profile becomes more peaked at the centre. Even if one rules out Q-stars
as models for pulsars (they have similar difficulties in this respect as
for strange stars) there still remains the possibility that they could be
an alternative model for the more massive objects in black-hole-candidate
systems.

The significance of the Q-star idea is not so much that it is widely
expected that the stellar-mass black hole candidates are really Q-stars
(although they might be), but rather that this stands as a concrete
counter-example to the normal discussion of the mass limit which it is
hard to rule out. It is true that there is a difference between the
predicted radii of Q-stars and black holes of similar mass and one might
hope to show that a particular object could not be a Q-star by picking up
a signal seeming to come from infalling matter at radii too small to be
outside the surface of a Q-star. However, it looks difficult to make this
distinction in practice because the difference in radius is so small.

\subsection{Searching for direct evidence of an event horizon}

So far, efforts to demonstrate the existence of stellar-mass black holes
have mostly been based on the calculated properties of neutron star models
(mainly the mass limit but now also the compactness) and this is, at best,
an indirect line of reasoning. The real point about black holes is the
existence of an event horizon with the properties anticipated from theory
and it is this that one would like to demonstrate in a clear way.
Calculation of neutron star models depends not only on our physical
description of high density matter at regimes which are not well-tested in
laboratory experiments but also, crucially, on the use of general
relativity to describe strong gravitational fields. If one used Newtonian
theory instead of general relativity, then all of the present stellar mass
black hole candidates could easily be accounted for as neutron stars even
with the most standard equations of state. While we do not believe that
Newtonian theory gives correct results under these circumstances, general
relativity is not the only possible alternative. Until rather recently,
this would have remained a serious uncertainty but now that results from
the binary pulsars [40] are placing very serious constraints on
alternative theories of gravity, the equation of state represents the
largest uncertainty.

What are the prospects for demonstrating the existence of an event horizon
in a more direct way? One line of reasoning which has attracted a lot of
attention in recent years is the following. What happens to the soft X-ray
transients when they are in quiescence? A model has been proposed by
Narayan and his colleagues (see [41] for a review) in which the accretion
rate is reduced as the SXT enters quiescence and the inner parts of the
disc swell up to form a structure which has been called an ADAF (Advection
Dominated Accretion Flow) [42]. This flow is very hot but of low density
and the energy liberated as material moves down the gravitational
potential well is mainly advected inwards with the infalling matter rather
than being emitted as radiation. Now, if the compact object is a black
hole, this energy simply passes through the event horizon and is lost to
view but if it is an object with a solid surface then the infalling matter
would be brought to rest at the surface and the excess energy would then
be emitted as radiation which could be observed. It is claimed that
observations of various sources (V404 Cygni in particular) match very well
with the picture of an ADAF around a black hole and that this is strong
evidence in favour of the presence of an event horizon. However, this
interpretation is still under debate and not everyone is convinced by it.

The coming of X-ray astronomy opened a new window onto the Universe for
the study of matter under extreme conditions near to compact objects. As
we have seen, it has brought us tantalizingly close to an unequivocal
demonstration that stellar mass black holes are really being observed but
the final step to being sure beyond reasonable doubt has proved to be very
difficult to make. Now, at the end of the twentieth century, we hope and
believe that we are on the threshold of opening another new window with
the detection of gravitational waves by laser interferometric detectors
and also, perhaps, by a new generation of bar detectors [43].
Gravitational waves can tell us directly about dynamical changes in
gravitational fields associated with very compact objects and bring us
information about the physical nature of the sources which cannot be
obtained in any other way. In detail, black holes {\it are} supposed to be
very different from any sort of compact star and the sort of additional
information which gravitational wave astronomy could provide may allow us
to see these differences more directly. 

\section{Super-massive black holes}

Following more than three decades of gradual increase in the number of
independent pieces of evidence indicating the existence of super-massive
black holes in the centres of some galaxies, there has been a remarkable
strengthening of the observational evidence in the last few years which
has produced a rather compelling case for the presence of dark massive
objects in the nuclei of most - if not all - galactic nuclei. Ironically,
black holes have now become perhaps the {\it least} exotic candidates for
identification with these dark objects.

Such powerful and robust evidence has been obtained chiefly thanks to
technological advances in observational techniques and instrumental
performance in the different energy bands, leading both to
improvements in the accuracy of measurements made with already
established observational methods and to the discovery of new - and
occasionally unexpected - pieces of information. The improvement of 
accuracy and sophistication in the modelling and interpretation of
data has, of course, been equally crucial.

\subsection{Collecting pieces of evidence}

The suggestion that super-massive black holes are hosted in the nuclei of
some galaxies originated at the beginning of the 1960s following the
discovery of quasars: the conversion into radiation of the gravitational
energy of matter in the potential well of such compact objects appeared to
be the most efficient way of producing the huge observed powers from
within small volumes.

Several phenomena associated with quasars and, more generally, with
powerful sources in the central cores of about 10\% of galaxies -
globally referred to as Active Galactic Nuclei (AGNs) - have since
contributed to reinforcing the belief that black holes are ultimately
responsible for the production of energy and activity in these galactic
cores (although alternative and more ``conventional'' hypotheses have also
been proposed).

The main characteristic feature of the AGN phenomenon is the inferred
compactness of the sources: luminosities of the order of $10^{46}$ erg
s$^{-1}$ (more than $10^{12}$ times the luminosity of the sun) are
produced from regions less than a light year across ($\sim 10^{18}$ cm).
(If the energy is taken to be emitted isotropically, the inferred
luminosity goes up to $10^{48}$ erg s$^{-1}$.) The most extreme constraint
on the compactness comes from the high energy (X-ray) radiation. Up to
several per cent of the total power can be emitted in this spectral band
with the radiation being highly variable on timescales of less than an
hour in some cases, thus setting extremely tight upper limits for the
typical dimensions of the region in which the energy is generated. This
high energy radiation, together with other spectral characteristics,
including line emission from gas moving at speeds of thousands of km
s$^{-1}$, cannot be satisfactorily ascribed to any stellar-related
(quasi-thermal) process.

There are two main facts which suggest that the AGN power source is
associated with a relativistic potential. Firstly, efficiencies of
mass-to-light conversion of the order of 10\% are required in order to
satisfy the observational constraints on both the amount of available fuel
in the host galaxies and the expected residual masses. These efficiences
greatly exceed that of nuclear fusion ($\sim$ 0.8\%). A black hole, as a
single super-massive object, is the most likely cause of the required deep
potential well. The tight constraints on the size of the volume involved,
the long-term stability of such central masses, the quite coherent
structure and lack of periodicities in the observed variability, all argue
against it being produced by clustering of ``small'' bodies. These
arguments strongly favour the case for the observed luminosity originating
from the extraction of gravitational energy from matter in the potential
wells of black holes with masses in the range $\sim$ 10$^6$--10$^{10}$
M$_{\odot}$ (see [44] and references therein).

Secondly, around 10\% of AGNs appear to be associated with the
presence of collimated structures (jets) along which plasma (presumably
ejected from the inner nucleus) sometimes moves at relativistic speeds,
corresponding to bulk Lorentz factors of the order of $10$. The associated
bulk kinetic power (and possibly also the magnetic power) can be of the
same order as the total luminosities. These structures can extend for up
to a million light years (i.e. through about ten orders of magnitude
increase in dimension) while remaining reasonably aligned. The jet
phenomenon thus requires the presence of a relativistic potential well and
a stable preferred axis over long timescales, properties which can
consistently be ascribed to the presence of a (spinning?) black hole at
the centre of the host galaxy (see [45] and references therein). The jet
power might even be extracted directly from the rotational spin energy
of
the hole itself by means of the Blandford-Znajek mechanism [46].

A further fact supporting the black hole conjecture is that, despite
significant phenomenological differences, the fundamental properties of
AGNs seem to be remarkably similar over a luminosity range of more than
six orders of magnitude (or even greater if one considers the analogy with
stellar-mass black hole candidates which we discuss in the Conclusion).

\subsection{Black hole demography}

In recent years, evidence has accumulated for the presence of massive
compact objects both in active galaxies (i.e. ones showing signs of
activity in the core) and (less predictably and more importantly) in
non-active galaxies, for which observations do not reveal any clear sign
of non-stellar processes taking place.

The crucial information which provides the clearest signature for not only
a central relativistic potential but, indeed, for a super-massive black
hole, are measures of (or limits on) the presence of a large, dark and
stable mass concentration within a small enough radius, such that the
inferred gravitational potential cannot easily be ascribed to aggregates
(dense clusters) of stars, black holes, brown dwarfs, planets or
elementary particles. Constraints on these alternatives come from the
requirement that the timescale for cluster evaporation and/or collapse
into a single object (depending on the mass, size and density of the
clustering objects) should be longer than the age of the system. In at
least two galaxies (the Milky Way and NGC 4258) the limits on the size are
so tight that the identification of the dark objects with super-massive
black holes seems to be almost unavoidable. 

The recent indications of the presence of dark massive objects in the
centres of galaxies are based on measurements of their effects on the
dynamics of stars and/or gas in the inner galactic core, within the
comparatively small radius where the potential of the central dark object
is dominant ($r \sim GM/\sigma^2 \sim 40 M_8 \sigma^{-2}_{100}$ parsecs,
where $M=10^8 M_8 M_{\odot}$ is the mass of the central object and
$\sigma=100 \sigma_{100}$ km s$^{-1}$ is the root mean square orbital
speed). Even for the (nearby) Virgo cluster, these observations require an
angular resolution better than $\sim 0.2$ seconds of arc. Stars and clouds
of gas within this radius have been considered as quasi test particles
orbiting a central mass with velocities given by $v^2 = \alpha G M /r$
where, however, $\alpha$ might need to be determined by detailed modelling
for the potential as being due to a point source together with the
extended galaxy and with the velocity distribution being anisotropic.

Use of both stellar and gas dynamics provides interesting advantages and
simplifications for the modelling and hence for the robustness of the
inferred results (see, for example, [47]).

The motion of stars is directly (and almost solely) affected by the
potential well of the galaxy and the central mass concentration: stars
thus behave basically like point masses in ballistic motion. However, the
random-motion component of their speed (only velocity components along the
line of sight can be measured) can exceed the bulk-motion one. The
velocity field can be significantly anisotropic since encounters between
stars are negligible and the relaxation time for the stellar system
exceeds the other relevant timescales and, in fact, suitable stellar
distributions are known to mimic the presence of a central massive object.
Therefore the results obtained from stellar dynamics are significantly
dependent on modelling (which, however, is helped by the fact that the
point-like mass potential can be considered as stationary, since the
growth timescale for the central mass is much longer than the orbital
timescale for the stars).

Gas dynamics, on the other hand, can in principle be influenced by forces
other than gravity (e.g. radiation pressure, pressure gradients, etc.).
However, since internal energy can quite easily be dissipated whereas
angular momentum cannot, the gas might plausibly be expected to be
relatively cold and in a disc-like structure, with the bulk-motion
component dominating the dynamics, thus simplifying difficulties
associated with the de-projection of the velocity field. Gas dynamical
measures have thus provided, in some cases, less ambiguous evidence for
the presence of a central mass and more secure measurements of how much
matter is there.

\subsection{Stellar dynamics}

Thanks to the unprecedentedly high angular resolution and sensitivity of
the Hubble Space Telescope (HST), it has been possible to measure the
spatial distribution and spectroscopic velocities of stars in the cores of
several nearby galaxies. Up to now, $\sim$ 15 robust mass determinations
have been made plus another $\sim$ 20 which are more critically model
dependent and these almost invariably point towards the presence of a
massive dark object in the centre of the galaxy (i.e. with corresponding
mass-to-light ratios greatly exceeding that for the sun), with the
inferred masses ranging between $\sim 10^6$ and a few$\times 10^9$
M$_{\odot}$ (see [48] and references therein). Among these is the case of
the nearby Andromeda galaxy (M31).

Three-dimensional stellar dynamics has also provided the strongest
evidence supporting the presence of a (mildly active) black hole in the
centre of our own Galaxy (as indeed predicted by Lynden--Bell \& Rees in
1971! [49]). Previous radio and infrared measurements of the gas motion
had shown that this was not determined only by the central gravitational
potential and thus proved inadequate for reaching a firm conclusion. The
new evidence had to be obtained using observations in the near infrared
band (performed with the New Technology Telescope at ESO) since our line
of sight towards the Galactic Centre is strongly obscured at optical and
ultraviolet wavelengths by the presence of dust and gas in the plane of
the Galaxy. The motion of individual stars in the central 0.01 parsec of
the Galaxy (with speeds exceeding thousands of km s$^{-1}$) was accurately
monitored over a period of a few years, providing measures of the
transverse component which, together with spectroscopic radial velocity
information, have allowed a remarkably accurate estimate to be made of the
central mass: $2.6\times 10^6$ M$_{\odot}$ (Eckart \& Genzel 1998 [50]).

\subsection{Gas discs}

Determinations of the dynamics of gas discs (and a few stellar discs)
within the inner $\sim 100$ parsecs of some nearby galaxies have been
obtained using HST observations. For several systems, photometric images
of (often dusty) discs have been obtained. In some cases, the emission of
ionized gas in certain spectral lines is sufficiently bright that
unambiguous evidence has been found for Keplerian orbital motion of a thin
distribution of gas (quasi-flat and therefore relatively cold) indicating,
a posteriori, that the gas dynamics is mainly subject to the gravitational
field produced by a central mass.

A remarkable example is the case of M87, in the Virgo Cluster, which shows
signs of low-power activity (the most conspicuous being the presence of
jets observed in the radio, optical and X-ray bands). Being at such low
redshift, a stellar cusp in the giant elliptical host galaxy had long been
observed [51] and a central mass of $\sim$ 10$^9 M_{\odot}$ had been
inferred but it was shown that the dynamics could also be accounted for by
an anisotropic velocity field. However, the presence of a central mass has
now become compelling following the determination of the velocity field of
a circumnuclear disc of ionized gas emitting in H$\alpha$, whose line
profiles and corresponding rotation curve are extremely consistent with
the gas being in Keplerian motion in a plane approximately perpendicular
to the jet direction. The measurements made imply the presence of a total
mass of $\sim 3.2 \times 10^9 M_{\odot}$ within the central 3.5 parsecs of
the galaxy (Macchetto et al 1997 [52] and references therein).

\subsection{Water maser discs}

An independent observational technique, which has provided one of the
strongest cases for the presence of super-massive black holes, is the
measurement of gas dynamics by means of the maser-emission line of water
at the wavelength of 1.3 cm.  Although several maser-emitting discs have
now been discovered, the most convincing and robust case remains that of
the spiral galaxy NGC 4258, perhaps because its emitting disc is observed
edge-on. The nearby spiral galaxy NGC 1068 provides the second best
example.

Radio measurements with the VLBA (Very Long Baseline Array) can achieve
angular resolutions $\sim$ 100 times better than HST (less than half of a
milliarcsecond) and show the presence of small H$_2$O masers distributed
over $\sim$ 0.1 parsec, confined within a single (although warped) plane.
Also, the VLBA spectral resolution is so accurate that the rotational
velocity of the molecular gas can be determined as being Keplerian motion
on circular orbits with a precision of better than 1\%. The inner edge of
the detected disc is at a radial distance of 0.13 parsec. The inferred
nuclear mass of $3.6\times 10^7 \,M_{\odot}$ confined within such a small
region is inconsistent with being a stable cluster of stars, strongly
pointing towards the presence of a black hole (Miyoshi et al. 1995 [53]).

\subsection{X--ray lines}

The measurements described so far are limited to mapping the dynamics of
stars and gas at distances larger than $\sim 10^4$ times the typical
radius of the inferred central black hole. Therefore, while the
observations provide some powerful evidence for the presence of massive
dark objects even in non-active galaxies, the dynamics of these systems
can be well-described in the Newtonian limit and do not provide any direct
test for general-relativistic effects. 

Most of the phenomenology associated with AGNs only requires the presence
of a deep gravitational potential well in which mass-energy can be
converted and radiated with high efficiency. Indeed, in Active Nuclei the
thermal and ionization properties of gas located at distances larger than
light days from the primary luminosity source are those relevant for the
emission of radiation at optical and longer wavelengths. However, higher
energy emission (typically X-rays) is produced at smaller radii, less than
a light hour away from the central engine, where the material is at a
higher temperature and the form of the potential well can imprint
interesting signatures on the emitted radiation. Measurements with high
sensitivity and high energy resolution of the X-ray spectra of nearby AGNs
have provided the first direct indication of material accreting in a
relativistic potential, as well as information about its spatial
distribution and accretion properties.

In particular, in the last decade observations have shown the relatively
common presence of emission lines in the X-ray band (mostly for nearby
Seyfert galaxies) at around 6-7 keV. These features and the shape of the
associated broad-band spectrum appear to be fully consistent with a
scenario in which a primary X-ray continuum source illuminates the surface
of a medium which is optically thick to Thomson scattering and then is
partially absorbed, partially Compton reflected and partially re-emitted
as line radiation. The strongest line is consistently accounted for as
being emission from the K$\alpha$ fluorescent transition of highly
(photo--)ionized iron atoms, which are sufficiently abundant to produce
the detected photon flux. The solid angle subtended by the reflecting gas
and its relatively low temperature ($\lap 10^6$ K) also suggest that this
can be consistently identified with matter accreting onto a central black
hole in the core of an AGN in a probably quasi-Keplerian thin disc.

The profile of a spectral line emitted from the inner parts of the disc
can provide information about the gravitational field at the location of
the radiating gas (Fabian et al. 1989 [54]). Recent exciting results for
such reprocessed spectra followed observations by the ASCA and BeppoSAX
X-ray satellites: the sensitivity and spectral resolution provided by the
on-board instruments allowed the profile of the iron line to be resolved,
convincingly revealing broad and asymmetric (skewed) features with
strongly redshifted tails (Tanaka et al 1995 [55]). The line widths imply
velocities of the order of a few times 10$^4$ km s$^{-1}$, thus suggesting
the motion of gas in a strong gravitational potential (alternative
hypotheses appear to be unsatisfactory). Furthermore, the line shape can
be best accounted for by the predicted effects of Doppler shifts and
gravitational redshift. Typical inferred radii for the emitting gas are
between a few and about ten times the Schwarzschild radius. In the best
studied source, MCG--6--30--15, the red side of the line extends down to
such low energies that the inner radius for the emitting gas distribution
is set at around $3r_s$. The central mass cannot be directly determined
from these measurements; however attempts have been made to infer a
black hole spin parameter from data modelling. Although this modelling is
probably currently too simplified (more than one plausible scenario has
been proposed) and the data are not of high enough quality to derive
robust estimates, X-ray spectroscopy does constitute a powerful and
promising astrophysical tool to probe the strong gravitational field
associated with the nuclei of galaxies and to determine the space-time
metric (see [56] and references therein).

\subsection{Some implications} 

The phenomenological evidence discussed so far for the existence of
super-massive black holes in the centres of galaxies has exciting
astrophysical implications and has led to intense theoretical work, both
from the point of view of their associated physics (``how do they work?'')
and of their cosmological role (``how do they form and evolve?'').

Despite the remaining large uncertainties (mostly related to the data
interpretation and lack of statistically complete samples of galaxies) the
inferred black hole masses and number density are quantitatively
consistent with the requirements imposed by the issue of how they are 
fuelled and by quasar evolution.

Here, we briefly mention some of the cosmological aspects related to
the recent findings which are currently being debated. Several of the most
crucial and recently re-stimulated issues are directly related to the
inference that there may be large black holes in the nuclei of most (if
not all) nearby galaxies. Interestingly, and rather independently, it has
been found that about 50\% of nearby galaxies might show some (low level)
nuclear activity. Also, the determination of black hole masses has
revealed that these are somehow related (with a rather large dispersion
and possibly subject to observational selection effects) to the mass of
the bulge (or spheroidal) component of the host galaxies, with the
inferred black hole masses being a few tenths of a percent of the bulge
masses. It appears that the existence and/or the formation of bulges of
galaxies is closely related to the central mass concentration in black
holes. Questions then arise not only about when and how super-massive
black holes were formed, but also (closely connected with this) about how
their formation is related to that of the whole host galaxy [57]. 

Since luminous quasars appear to be present already at cosmological
redshift $z \sim 5$, strong requirements are set on the rapidity of the
initial black hole formation. Although this might have begun with the
collapse or merger of initial seeds of the order of tens of solar masses
which then subsequently grew through accretion up to typical masses of
$10^8 M_{\odot}$, probably the most widely-accepted view associates the
formation of black holes with the initial collapse of gas possibly left
over from the same cloud from which stars initially condensed [58].
Following the initial collapse, the main luminous quasar phase would then
take place, with black hole masses growing to $\sim 10^6-10^9 M_{\odot}$.
Later activity, corresponding to the most recent AGN phenomena,
stimulated for example by mergers and close galaxy encounters, would
then constitute a relatively minor event.

\section{Conclusion}

In this article we have described the situation regarding astrophysical
evidence for the existence of black holes in two distinct mass ranges
corresponding to stellar masses and the masses typical of galactic nuclei.
Evidence has been progressively mounting and the case is now rather strong
for saying that black holes have indeed been observed, particularly on the
super-massive scale where there now seems little alternative to accepting
this conclusion in the case of the best candidates. For stellar-mass
candidates, the case already looks conclusive if one accepts the
conventional view of the maximum-mass limit for sufficiently-compact
alternatives to black holes. However, the argument leading to this mass
limit still leaves some possible scope for doubt.

While the discussion of the properties of black hole candidates is
rather different for the two mass ranges, there are many similarities
between the two scales which may be usefully exploited. In particular,
great interest has been aroused over the last few years by observations of
so-called ``micro-quasars'' [59] (of which the best current example is the
source GRS 1915+105). These are stellar-mass black hole candidate systems
which show many of the properties of AGNs (including expulsion of material
along the rotation axis) but with characteristically shorter timescales
(roughly proportional to the inferred black-hole mass). Because these
objects are easier to observe than AGNs (both because they are closer and
also because of the shorter timescales involved), this link between the
different scales may provide important new insights into how AGNs work.

During the preparation of this article, it was announced that some
$\alpha$-process elements (oxygen, magnesium, silicon and sulphur) have
been detected in the normal companion star of the binary black-hole
candidate Nova Scorpii [60] (thought to be a black hole of mass $5.5 - 7.9
M_\odot$ [61]). These cannot have originated in the normal star and the
most obvious explanation is that they were transferred to it as a result
of a supernova explosion of the progenitor of the black hole candidate,
thus giving observational confirmation for previous theoretical results
suggesting that black holes could result from supernova explosions.

Study of iron-line profiles (as discussed in Section 3) is potentially a
powerful diagnostic tool for determining the properties of black hole
candidates although it is much easier to observe them for the
super-massive objects than for the stellar-mass ones where the strength of
the iron lines with respect to the continuum is much weaker. X-ray
observations with various newly-launched satellites (e.g. Chandra) and
future planned missions (XMM, Astro-E but more powerfully Constellation X)
will provide a major impetus for this work particularly in connection with
AGNs. Other effects connected with black holes which could be fruitful
objectives for future observational study (see the review by Rees [58])
include quasi-periodic oscillations of accretion discs around black holes,
in both mass ranges, and optical (and possibly X-ray) flashes resulting
from the occasional capture and tidal disruption of stars by super-massive
black holes. Finally, we hope that gravitational wave astronomy, which is
the subject of another article in this volume [43], will soon provide a
new way of looking at both stellar-mass candidates (using earth-bound
detectors) and super-massive ones (probably needing to wait for detectors
in space).

\numreferences

\numrefjl {[1]}{Michell J 1784}{Phil. Trans. R. Soc. (London)}{74}{35}

\numrefbk {[2]}{Laplace P S 1796}{Exposition du Syst\`eme du Monde}
{(Paris), vol. 2, p. 305}

\numrefbk {[3]}{Israel W 1987}{in Three Hundred Years of Gravitation}{eds.
S W Hawking, W Israel, Cambridge University Press, p. 199}

\numrefjl {[4]}{Schwarzschild K 1916} {Preuss. Akad. Wiss. Berlin,
Sitzber.} {1916}{189}

\numrefjl {[5]}{Einstein A 1915} {Preuss. Akad. Wiss. Berlin, Sitzber.}
{1915}{844}

\numrefjl {[6]}{Lodge O 1921}{Phil. Mag.}{41}{549}

\numrefjl {[7]}{Chandrasekhar S 1931}{Astrophys. J.}{74}{81}

\numrefjl {[8]}{Eddington A S 1935}{Observatory}{58}{37}

\numrefjl {[9]}{Lemaitre G 1933}{Ann. Soc. Sci. (Bruxelles)}{A53}{51}

\numrefjl {[10]}{Einstein A 1939}{Ann. Math. (Princeton)}{40}{922}

\numrefjl {[11]}{Oppenheimer J R and Snyder H 1939}{\PR}{56}{455}

\numrefbk {[12]}{Bergmann P G 1942}{An Introduction to the Theory of
Relativity}{Prentice-Hall: New York, pp. 203-4}

\numrefjl {[13]}{Kruskal M D 1960}{\PR}{119}{1743}

\numrefbk {[14]}{Harrison B K, Thorne K S, Wakano M and Wheeler J A 1965}
{Gravitation Theory and Gravitational Collapse}{University of Chicago
Press}

\numrefjl {[15]}{Kerr R P 1963}{\PRL}{11}{237}

\numrefjl {[16]}{Israel W 1967}{\PR}{164}{1776}

\numrefbk {[17]}{Carter B 1973}{in Black Holes}{eds. C DeWitt and B.S.
DeWitt, Gordon \& Breach, New York, p. 57}

\numrefjl {}{Hawking S W 1972}{Commun. Math. Phys.}{33}{323}

\numrefjl {}{Robinson D C 1975}{\PRL}{34}{905}

\numrefjl {[18]}{Penrose R 1965}{\PRL}{14}{57} 

\numrefjl {[19]}{Hawking S W 1966}{\PRS}{A294}{511}

\numrefjl {[20]}{Geroch R P 1966}{\PRL}{17}{445}

\numrefjl {[21]}{Hawking S W 1974}{Nature}{248}{30}

\numrefjl {[22]}{Jedamzik K 1998}{Phys. Rept.}{307}{155}

\numrefbk {[23]}{Carr B J 1993}{in The Renaissance of General Relativity
and Cosmology}{eds. G F R Ellis, A Lanza and J C Miller, Cambridge
University Press, p. 258}

\numrefbk {[24]}{Blandford R D 1987}{in Three Hundred Years of
Gravitation}{eds. S W Hawking, W Israel, Cambridge University Press, p.
277}

\numrefjl {[25]}{Giacconi R H, Gursky H, Paolini F R and Rossi B B 1962}
{\PRL}{9}{439}

\numrefjl {[26]}{Shklovskii I S 1967}{Astron. Zhur.}{44}{930}

\numrefjl {[27]}{Prendergast K H and Burbidge G R 1968}{Astrophys. J.
Lett.}{151}{L83}

\numrefjl {[28]}{Bradt H V D and McClintock J E 1983}{Ann. Rev. Astron.
Astrophys.}{21}{13}

\numrefjl {[29]}{Rothschild R E E, Boldt E A, Holt S S and Serlemitsos P J
1974}{Astrophys. J. Lett.}{189}{L13}

\numrefjl {[30]}{Gies D R and Bolton C T 1986}{Astrophys. J.}{304}{371}

\numrefjl {[31]}{Bahcall J N, Dyson F J, Katz J I and Paczynski B 1974}
{Astrophys. J. Lett.}{189}{L17}

\numrefjl {}{Fabian A C, Pringle J R and Whelan J A J 1974}{Nature}{247}
{351}

\numrefbk {[32]}{Charles P A 1998}{in Theory of Black Hole Accretion
Disks}{eds. M A Abramowicz, G Bjornsson and J E Pringle, Cambridge
University Press, p. 1}

\numrefjl {[33]}{Shahbaz T, Naylor T and Charles P A 1993}{MNRAS}{265}
{655}

\numrefjl {}{Shahbaz T, Naylor T and Charles P A 1994}{MNRAS}{268}{756}

\numrefjl {}{Shahbaz T, Ringwald F A, Bunn J C, Naylor T, Charles P A and
Casares J 1994}{MNRAS}{271}{L10}

\numrefjl {[34]}{Rhoades C E and Ruffini R 1974}{\PRL}{32}{324}

\numrefjl {[35]}{Hartle J B 1978}{Phys. Repts.}{46}{201}

\numrefjl {[36]}{Friedman J L and Ipser J R 1987}{Astrophys. J.}{314}{594}

\numrefjl {[37]}{Witten E 1984}{\PR}{D30}{272}

\numrefjl {}{Haensel P, Zdunik J L and Schaeffer R 1986}{Astron.
Astrophys.}{160}{121}

\numrefjl {}{Alcock C, Farhi E and Olinto A 1986}{Astrophys. J.}{310}{261}

\numrefjl {[38]}{Bahcall S, Lynn B W and Selipsky S L 1989}{\NP}{B325}
{606}

\numrefjl {}{Bahcall S, Lynn B W and Selipsky S L 1990}{Astrophys. J.}
{362}{251}

\numrefjl {[39]}{Miller J C, Shahbaz T and Nolan L A 1998}{MNRAS}{294}
{L25}

\numrefjl {[40]}{Stairs I H, Arzoumanian Z, Camilo F, Lyne A G, Nice D J,
Taylor J H, Thorsett S E and Wolszczan A 1998}{Astrophys. J.}{505}{352}

\numrefbk {[41]}{Menou K, Quataert E and Narayan R 1999} {in Black Holes,
Gravitational Radiation, and the Universe}{eds. B R Iyer and B Bhawal,
Kluwer, p. 265}

\numrefjl {[42]}{Narayan R and Yi I 1994}{Astrophys. J.}{428}{L13}

\numrefjl {}{Abramowicz M A, Chen X, Kato S, Lasota J-P and Regev O 1995}
{Astrophys. J.}{438}{L37}

\numrefjl {[43]}{Schutz B F 1999}{\CQG}{(this volume)}{}

\numrefjl {[44]}{Rees M J 1984}{Ann Rev Astron Astrophys}{22}{471}

\numrefbk {}{Blandford R D 1990}{Active Galactic Nuclei -- $20^{th}$
SAAS-FEE Course}{eds. T J-L Courvoisier and M Mayor, Springer-Verlag}

\numrefbk {}{Krolik J H 1999}{Active Galactic Nuclei: From the Central
Black Hole to the Galactic Environment}{Princeton Series in
Astrophysics, Princeton University Press}

\numrefjl {[45]}{Begelman M C, Blandford R D and Rees M J 1984}
{\RMP}{56}{255}

\numrefjl {[46]}{Blandford R D and Znajek R L 1977}{MNRAS}{179}{433}

\numrefjl {[47]}{Kormendy J and Richstone D O 1995}{Ann. Rev. Astron.
Astrophys.} {33}{581}

\numrefjl {[48]}{Richstone D O et al 1998}{Nature}{395}{14}

\numrefbk {}{van der Marel R 1998}{Proc. IAU Symp. 186: Galaxy
Interactions and High and Low redshifts}{eds B Sanders and J Barnes,
Kluwer, in press}

\numrefjl {[49]}{Lynden-Bell D and Rees M J 1971}{MNRAS}{152}{461}

\numrefbk {[50]}{Eckart A and Genzel R 1998}{The Central Parsecs}
{eds. H Falcke, A Cotera, W Huschl, F Melia and M Rieke, ASP 
Conference Series, in press}

\numrefjl {[51]}{Sargent W L W, Young P J, Lynds C R, Boksenberg A,
Shortridge K and Hartwick F 1978}{Astrophys. J.}{221}{731}

\numrefjl {[52]}{Macchetto F, Marconi A, Axon D J, Capetti A, Sparks W and
Crane P 1997}{Astrophys. J.}{489}{579}

\numrefjl {[53]}{Miyoshi M, Moran J, Herrnstein J, Greenhill L, Nakai N,
Diamond P and Inoue M 1995}{Nature}{373}{127}

\numrefjl {[54]}{Fabian A C, Rees M J, Stella L and White N E 1989}{MNRAS}
{238}{729}

\numrefjl {[55]}{Tanaka Y et al 1995}{Nature}{375}{659}

\numrefbk {[56]}{Fabian A C 1998}{Eighth Astrophysical Conference on
Accretion Processes in Astrophysical Systems: Some Like It Hot}{eds. S
Holt and T R Kallman, AIP Conf. Proc. 431, 246}

\numrefbk {[57]}{Blandford R D 1999}{to appear in ASP Conference Series}
{eds. D Meritt, M Valluri and J Sellwood [astro-ph/9906025]}

\numrefjl {}{Salucci P, Szuszkiewicz E, Monaco P and Danese L 1999}{MNRAS}
{307}{637}

\numrefbk {[58]}{Rees M J 1999}{to appear in Black Holes and Relativity -
Proceedings of Chandrasekhar Memorial Conference, Chicago, Dec. 1996} {ed.
R M Wald [astro-ph/9701161]}

\numrefjl {[59]}{Mirabel I F and Rodriguez L F 1999}{Ann. Rev. Astron.
Astrophys}{37}{}

\numrefjl {[60]}{Israelian G, Rebolo R, Basri G, Casares J and Martin E L
1999}{Nature}{401}{142}

\numrefjl {[61]}{Shahbaz T, van der Hooft F, Casares J, Charles P A and
van Paradijs J 1999}{MNRAS}{306}{89}

\ack

We gratefully acknowledge many helpful discussions with colleagues in
Oxford, Cambridge, Trieste and elsewhere. Financial support for our work
is provided by the Italian Ministero dell'Universit\`a e della Ricerca
Scientifica e Tecnologica.

\vfill\eject\end